\documentclass[10pt,journal,compsoc]{static/package/IEEEtran}
\usepackage{diagbox,amsmath,amsthm,amssymb,booktabs,comment,graphicx,subfigure,multirow,bm,tikz}
\usepackage{tabularx,array,float,colortbl,lstautogobble,color,zi4,bbm}
\usepackage[ruled,linesnumbered]{algorithm2e}
\usepackage[T1]{fontenc}
\usepackage[symbol]{footmisc}
\usepackage{algpseudocode}
\usepackage{tabularx}
\usepackage{amssymb}
\usepackage{tabularx,multirow,colortbl}
\usepackage{array}
\usepackage{float}
\usepackage{verbatim}
\usepackage{diagbox}
\usepackage{utfsym}
\usepackage{marvosym}

\def\checkx{\scalebox{0.75}{\usym{2613}}}
\def\checky{\checkmark}

\newtheorem{theorem}{Theorem}[section]

\theoremstyle{definition}
\newtheorem{definition}{Definition}[section]
\newtheorem{proposition}[theorem]{Proposition}
\ifCLASSOPTIONcompsoc
  \usepackage[nocompress]{cite}
\else
  \usepackage{cite}
\fi

\ifCLASSINFOpdf
\else
\fi
\begin{document}
\title{\textsc{FDINet}: Protecting against DNN Model Extraction using Feature Distortion Index}
\author{
  Hongwei~Yao,
  Zheng~Li,
  Haiqin~Weng,
  Feng~Xue,
  \and
  Zhan~Qin~{\Letter},
  Kui~Ren \IEEEmembership{Fellow, IEEE}
  \IEEEcompsocitemizethanks{
    \IEEEcompsocthanksitem Kui~Ren, Zhan~Qin and Feng Xue are with The State Key Laboratory of Blockchain and Data Security, Zhejiang University, Hangzhou, China.
    \textbf{Zhan~Qin is the corresponding author}.
    (E-mail: qinzhan@zju.edu.cn).
    \IEEEcompsocthanksitem Zheng Li is with the German National Big Science Institution within the Helmholtz Association, Saarbrücken, German (E-mail:zheng.li@cispa.de).
    \IEEEcompsocthanksitem Hongwei~Yao is with The State Key Laboratory of Blockchain and Data Security, Zhejiang University and Department of Computer Science City University of Hong Kong.
    (E-mail: yhongwei@zju.edu.cn).
  }
}
\markboth{IEEE TRANSACTIONS ON DEPENDABLE AND SECURE COMPUTING, VOL. 1, NO. 1, JUNE~2023}%
{Shell \MakeLowercase{\textit{et al.}}: Bare Advanced Demo of IEEEtran.cls for IEEE Computer Society Journals}

\IEEEtitleabstractindextext{
\begin{abstract}
  Machine Learning as a Service (MLaaS) platforms have gained popularity due to their accessibility, cost-efficiency, scalability, and rapid development capabilities.
  However, recent research has highlighted the vulnerability of cloud-based models in MLaaS to model extraction attacks.
  In this paper, we introduce \textsc{FDINet}, a novel defense mechanism that leverages the feature distribution of deep neural network (DNN) models.
  Concretely, by analyzing the feature distribution from the adversary's queries, we reveal that the feature distribution of these queries deviates from that of the model's problem domain.
  Based on this key observation, we propose Feature Distortion Index (FDI), a metric designed to quantitatively measure the feature distribution deviation of received queries.
  The proposed \textsc{FDINet} utilizes FDI to train a binary detector and exploits FDI similarity to identify colluding adversaries from distributed extraction attacks.
  We conduct extensive experiments to evaluate \textsc{FDINet} 
    against six state-of-the-art extraction attacks on four benchmark datasets and four popular model architectures.
  Empirical results demonstrate the following findings:
  (1) \textsc{FDINet} proves to be highly effective in detecting model extraction, achieving a \textbf{100\% detection accuracy} on DFME and DaST.
  (2) \textsc{FDINet} is highly efficient, using just 50 queries to raise an extraction alarm with an \textbf{average confidence of 96.08\%} for GTSRB.
  (3) \textsc{FDINet} exhibits the capability to identify colluding adversaries with an accuracy \textbf{exceeding 91\%}.
  Additionally, it demonstrates the ability to detect two types of adaptive attacks.
\end{abstract}
\begin{IEEEkeywords}
Model extraction, model stealing, Feature Distortion Index.
\end{IEEEkeywords}
}
\maketitle
\IEEEdisplaynontitleabstractindextext
\IEEEpeerreviewmaketitle
\ifCLASSOPTIONcompsoc
\IEEEraisesectionheading{
  \section{Introduction}\label{sec:intro}
}
\else
\section{Introduction}
\label{sec:intro}
\fi

\IEEEPARstart{A}{s} the performance of deep neural networks (DNN) remarkably improves, 
DNN has been widely used in various fields (e.g., image recognition and natural language processing).
However, the construction of high-performance DNN models requires tremendous amounts of training data and computational resources, making it challenging for end-users to create their own private models.
Therefore, many companies choose to deploy DNN models to Cloud Service Providers (CSP) in order to offer online paid services through Machine Learning as a Service (MLaaS)~\cite{zhang2020mlmodelci}.
Recent reports even predict a remarkable economic boost of 21.72 billion dollars in the MLaaS market~\cite{machine2023market}.
Unfortunately, the value associated with these models also leads to the emergence of model extraction attacks (also known as model stealing attacks)~\cite{tramer2016stealing,jagielski2020high}.

In MLaaS, only the CSP has access to the parameters and architecture of the cloud-based model.
The clients can only interact with the model through a public API.
While the cloud-based model may appear as a black-box to clients, it is still possible for a malicious client to interact with the model and replicate its behavior using input-output pairs.
This poses a significant risk of privacy breach for the cloud-based models.
Recent studies~\cite{shen2022model,sha2022can,kalpesh2020thieves,gong2020model} have shown that an adversary can launch model extraction attacks by querying MLaaS, imitating the behaviors of the target DNN model, and creating a surrogate model.
Furthermore, using the extracted surrogate model, the adversary can launch additional attacks, including membership inference attacks~\cite{shokri2017membership}, adversarial examples~\cite{papernot2017practical,zhou2020dast,wang2021delving,ma2021simulating}, and model explanations~\cite{bastani2017interpreting,kazhdan2020meme}.
Consequently, the protection of cloud-based models against model extraction attacks emerges as a critical issue that demands increased attention.

To enhance the security of MLaaS, there have been growing research efforts on model extraction detection~\cite{kesarwani2018model,juuti2019prada,sadeghzadeh2021hardness,zhang2021seat}.
Existing detection approaches typically involve analyzing query distributions and utilizing statistical tools to identify potentially malicious queries.
Although the existing detection approaches have made promising progress, they still have several limitations.
One key limitation is that most existing methods rely on strong assumptions about the adversary, which limits their generalizability to different extraction attacks. 
For example, PRADA~\cite{juuti2019prada} is designed specifically to detect adversarial example-based queries, DeepDefense~\cite{lee2022model} fails to identify synthetic data-based queries.
As a result, it remains a challenge to identify the intrinsic characteristics of extraction attacks and develop a generic detection method to identify diverse attacks.
Second, existing detectors need to maintain local proxy models~\cite{kesarwani2018model}, historical queries~\cite{juuti2019prada}, or training points~\cite{sadeghzadeh2021hardness}.
While these components contribute to detection accuracy, they may fail to identify malicious clients efficiently. 
Therefore, improving the efficiency of detection methods poses a significant challenge.
Furthermore, advanced stealth attacks, such as distributed model extraction attacks, can evade the existing detectors.
To the best of our knowledge, there is still no countermeasure to defend against the distributed attack, which adopts multi-clients to launch the same model extraction attack.
Thus, devising an effective defense to mitigate the impact of advanced stealth attacks is a pressing issue.

To address the aforementioned limitations, we propose \textsc{FDINet}, a generic effective and efficient detector against model extraction attacks that can be easily integrated into MLaaS systems.
In order to identify the intrinsic differences between benign and malicious queries, we investigate the attack strategies.
Concretely, we analyze the queries submitted by adversaries and make a motivative observation: the feature distribution of the adversaries’ queries deviates from that of the training set.
We refer to it as ``feature distortion,'' which is a universal characteristic across various model extraction attacks.
Based on this observation, we introduce the Feature Distortion Index (FDI), a metric designed to quantitatively measure the feature distribution deviation of received queries.
Furthermore, we observe a high degree of FDI similarity among queries generated by the same model extraction strategy.
This observation opens up new possibilities for identifying colluding adversaries in distributed extraction attacks. 
Leveraging this insight, we propose a distributed extraction attack detector with the capability of identifying colluding adversaries.
Additionally, we consider the adaptive adversary who knows our defense strategy may correct features before submitting queries to MLaaS. 
We propose an adaptive attack, namely \textit{Feature Correction, FeatC}, to evade our defense.


To validate the efficacy of \textsc{FDINet}, we conduct extensive experiments using four benchmark datasets, namely CIFAR10, GTSRB, CelebA, and Skin Cancer Diagnosis.
The evaluation results demonstrate that the proposed \textsc{FDINet} effectively and efficiently reveals malicious clients and identifies colluding adversaries.
The major contributions of this paper are summarized as follows:
\begin{enumerate}
    \item \textbf{Proposal of Feature Distortion Index (FDI)}: 
    We propose a novel metric, the FDI, to measure the feature deviation of submitted queries.
    By utilizing the FDI, we train a binary detector that accurately identifies malicious queries.

    \item \textbf{Identifying colluding adversaries}:
    We propose a distributed model extraction attack in which the adversary controls multiple colluding clients to send queries with a common attack goal.
    We analyze the FDI similarity of queries and develop a novel classifier that can identify colluding adversaries.
    This classifier is a pioneering approach in defending against distributed model extraction attacks.

    \item \textbf{Proposal of the adaptive attack, \textit{FeatC}}:
    In order to assess the robustness of \textsc{FDINet}, we propose an adaptive attack called \textit{FeatC}, specifically designed to bypass our defense mechanism.
    
    \item \textbf{Extensive experiments and evaluations}:
    We conduct extensive experiments to evaluate the performance of \textsc{FDINet} on four benchmark datasets and four popular model architectures.
    The results demonstrate the effectiveness and efficiency of \textsc{FDINet} in detecting malicious queries.
    Additionally, our approach is robust that achieves high performance in identifying colluding adversaries and two types of adaptive attacks.
\end{enumerate}

\begin{table*}[!t]
\caption{Taxonomy of real-time model extraction defense approaches.}
\label{tab:taxonomy_defense}
\centering
\resizebox{0.95\linewidth}{!}{
    \begin{tabular}{l|c|cc|ccc|c|c|c}
    \specialrule{1pt}{2pt}{2pt}
    \multirow{2}{*}{\textbf{Method}} & \multirow{2}{*}{\textbf{Type}}
        & \multicolumn{2}{c|}{\textbf{Support Outputs}} & \multicolumn{3}{c|}{\textbf{Defending Capability}}
        & \multirow{2}{*}{\shortstack{\textbf{Dummy}\\\textbf{Query}}}
        & \multirow{2}{*}{\shortstack{\textbf{Feature-corrected}\\\textbf{Query (Ours)}}} 
        & \multirow{2}{*}{\shortstack{\textbf{Distributed}\\\textbf{Attack (Ours)}}} \\
    \cline{3-7}
    & & Hard-label & Soft-label & $\mathcal{A}_{sur}$ & $\mathcal{A}_{adv}$ & $\mathcal{A}_{syn}$ & & & \\
    \specialrule{1pt}{0pt}{0pt}
    Kariyappa~\textit{et al.}~\cite{kariyappa2020protecting}’2020 &Active&\checkx&\checky     &\checky&\checky&\checkx        &\checky&\checkx&\checkx\\
    Zheng~\cite{zheng2019bdpl,zheng2020protecting}’2022             &Active&\checkx&\checky     &\checky&\checkx&\checkx        &\checkx&\checkx&\checkx\\
    Tribhuvanesh~\textit{et al.}~\cite{Tribhuvanesh2020Prediction}’2020&Active&\checkx&\checky  &\checky&\checky&\checkx   &\checkx&\checkx&\checkx\\
    Adam~\textit{et al.}~\cite{adam2022increasing}’2022           &Active&\checkx&\checky     &\checky&\checky&\checky        &\checkx&\checkx&\checkx\\
    Kariyappa~\textit{et al.}~\cite{kariyappa2020defending}’2020  &Active&\checkx&\checky   &\checky&\checky&\checkx  &\checkx&\checkx&\checkx\\
    
    Kesarwani~\textit{et al.}~\cite{kesarwani2018model}’2018       &Passive&\checky&\checky    &\checky&\checky&\checkx    &\checkx&\checkx&\checkx\\
    Juuti~\textit{et al.}~\cite{juuti2019prada}’2019              &Passive&\checky&\checky    &\checkx&\checky&\checkx    &\checky&\checkx&\checkx\\
    Zhang~\textit{et al.}~\cite{zhang2021seat}’2021               &Passive&\checky&\checky    &\checkx&\checky&\checky    &\checkx&\checkx&\checkx\\
    Pal~\textit{et al.}~\cite{pal2021stateful}’2021               &Passive&\checky&\checky    &\checky&\checky&\checky    &\checkx&\checkx&\checkx\\
    Lee~\textit{et al.}~\cite{lee2022model}’2022                  &Passive&\checky&\checky    &\checky&\checkx&\checkx    &\checkx&\checkx&\checkx\\
    
    \textbf{\textsc{FDINet} (Ours)}                                &Passive&\checky&\checky    &\checky&\checky&\checky        &\checky&\checky&\checky\\
    \specialrule{1pt}{2pt}{2pt}
    \end{tabular}
}
\end{table*}

\section{Related Works}
\label{sec:related}
\subsection{Model Extraction Attacks}
\label{sec:related_atk}
The concept of model extraction and the demonstration of its feasibility of 
  stealing intellectual property from private models on commercial MLaaS was initially proposed by Tram{\`e}r \textit{et al}. ~\cite{tramer2016stealing}.
The main principle of model extraction is to replicate the behavior and functionality of 
  a black-box victim model by analyzing query submissions and their corresponding outputs.
In this context, the selection of representative data plays a crucial role in determining the efficiency of model extraction attacks.
According to the strategy of sample selection, extraction attacks can be categorized into:

\textbf{Surrogate data-based schemes ($\mathcal{A}_{sur}$).}
In this scenario, the adversary possesses a comprehensive surrogate dataset, such as ImageNet and Flickr, 
  which consists of both problem domain (PD) and non-problem domain (NPD) samples.
To enhance the efficiency of query selection, the adversary commonly employs active learning strategies
  (e.g., Knockoff~\cite{orekondy2019knockoff}, ActiveThief~\cite{pal2020activethief}, 
  PAC-based active extraction~\cite{chandrasekaran2020exploring}, CopycatCNN~\cite{correia2018copycat}, 
  Bert Stealing~\cite{kalpesh2020thieves}, and GNN Stealing~\cite{he2021stealing}).

\textbf{Adversarial example-based schemes ($\mathcal{A}_{adv}$).}
The adversary is assumed to have access to a limited number of the problem domain data. 
In this scenario, the adversary crafts adversarial examples using primitive data, intending to approximate the decision boundary of the target model
(e.g., Jacobian-based Augmentation (JBA)~\cite{papernot2017practical}, T-RND~\cite{juuti2019prada}, DualCF~\cite{wang2022dualcf}, and Cloud-Leak~\cite{yu2020cloudleak}).

\textbf{Synthetic data-based schemes ($\mathcal{A}_{syn}$).}
In this scenario, the adversary employs a generative model to craft large-scale synthetic samples.
For example, Black-box Ripper~\cite{barbalau2020black}, Data-free Substitute Training (DaST)~\cite{zhou2020dast}, 
Data-free Model Extraction (DFME)~\cite{truong2020data}, DFMS-HL~\cite{sanyal2022towards}, MEGEX~\cite{miura2021megex} and MAZE~\cite{kariyappa2021maze}.

\textbf{Hybrid schemes ($\mathcal{A}_{hyb}$).} 
The idea behind hybrid schemes is to improve the effectiveness and efficiency of the model extraction attack 
by combining the strengths of each type of attack and mitigating their limitations (e.g., InverseNet~\cite{xueluan2021inversenet}, DivTheft~\cite{ma2023divtheft}).

Since the $\mathcal{A}_{hyb}$ scenario is the combining of other scenarios,
  we focus on $\mathcal{A}_{sur}$, $\mathcal{A}_{adv}$, and $\mathcal{A}_{syn}$ adversaries in this study.
It should be noted that those attacks compass a wide range of cutting-edge techniques, covering diverse attack scenarios.

\subsection{Defenses against Model Extraction Attacks}
\label{sec:related_def}
The countermeasures against model extraction attacks can be categorized into real-time defense and post-stealing defense.
Real-time defense aims to detect and prevent the extraction process when the stealing action is in-progress.
On the other hand, post-stealing defense strategies utilize copyright verification techniques, 
  such as DNN Watermarking~\cite{szyller2021dawn,jia2021entangled}, 
  DNN Fingerprinting~\cite{lukas2019deep,chen2022teacher,pan2022metav}, 
  or Dataset Inference~\cite{pratyush2021dataset,dziedzic2022dataset,zhu2021defending,li2022defending},
  to verify the ownership of the potentially stolen model.

This paper focuses on real-time model extraction defense, which comprises two primary branches of techniques: passive defense and active defense schemes. 
To provide a comprehensive overview and comparison of the those defense methods, we present a taxonomy of these techniques in Table~\ref{tab:taxonomy_defense}.
In the following paragraphs, we will discuss those two branches of methods.

\textbf{Passive defense.}
The passive defense approach aims to detect and interrupt malicious actions 
by monitoring and analyzing the distribution (e.g., abnormal distributions, significant information gain) of incoming queries
~\cite{kesarwani2018model,juuti2019prada,zhang2021seat,pal2021stateful,lee2022model}.
PRADA~\cite{juuti2019prada} keeps track of the minimum L2-norm distance 
between last sample and previous samples for each client to detect the adversary.
Since PRADA is based on the assumption that samples submitted by an adversary deviate from a normal (Gaussian) distribution.
Consequently, it is unable to detect queries that conform to a normal distribution.
Extraction Monitor (EM)~\cite{kesarwani2018model} employs local proxy models to quantify the extraction status of an individual client.
The proxy model measures either the information gain or the coverage of the feature space spanned by queries to estimate the learning status of individual and colluding adversaries.
However, EM has two main drawbacks:
(1) employing local proxy models results in high memory consumption that degrades the efficiency of MLaaS,
(2) large false alarms may be raised for the benign client.

\textbf{Active defense.}
The active defense approach involves adding perturbation to the victim model’s output
~\cite{zheng2019bdpl,zheng2020protecting,quiring2018forgotten,lee2019defending,chen2020ast,Tribhuvanesh2020Prediction,kariyappa2020protecting,kariyappa2020defending,dziedzic2022increasing,mazeika2022steer}.
Orekondy \textit{et al.} propose Prediction Poisoning~\cite{Tribhuvanesh2020Prediction}, 
which adds utility-constrained perturbations on the posteriors.
The perturbations maximize the angular deviation between the gradient of the poisoned posteriors and that of the victim model.
Zheng \textit{et al.} propose BDPL~\cite{zheng2019bdpl,zheng2020protecting}, 
which exploits Differential Privacy to add obfuscating noises on the prediction logits.

Passive defense strategies are the main focus of this paper due to the inherent limitations of active defense methods.
Active defense approaches rely on strong assumptions about the output forms, which give rise to two significant drawbacks.
Firstly, introducing perturbed probabilities may negatively impact the utility of cloud-based models.
Secondly, these methods are not applicable in scenarios where hard-label outputs are used instead of probability vectors.
Therefore, this paper primarily discusses the utilization of passive defense strategies.

\section{Preliminaries}
\label{sec:problem_description}
A Deep Neural Network (DNN) model is a function ${F}: \mathcal{X} \rightarrow \mathcal{Y}$ 
    parameterized by a set of parameters, 
    where $\mathcal{X} \in \mathbb{R}^{d}$ denotes the $d$-dimensional feature space and $\mathcal{Y}$ represents the output space.
For an online MLaaS application, 
    the private DNN model ${F}_{V}$ is first trained by the developer
    using enormous training data $\mathcal{D}_{train}$ to achieve a high accuracy on testing set $\mathcal{D}_{test}$
    and then deployed to the CSP.
Through querying prediction API using a pay-as-you-go manner, 
    the client can access prediction probabilities ${F}_{V}\left(x\right)$ for any given input data $x$.
The goal of model extraction is to create a surrogate model ${F}_{S}$ that replicates the functionality of the black-box victim model ${F}_{V}$.

\begin{figure*}[!ht]
  \centering
  \includegraphics[width=0.85\linewidth]{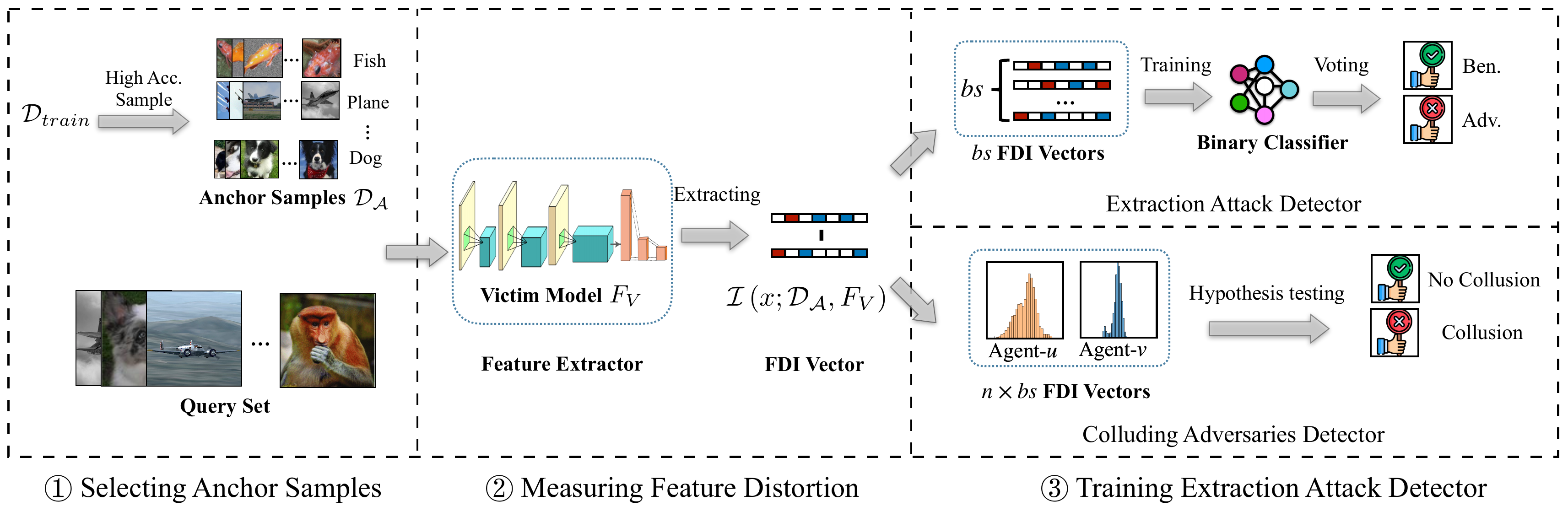}
  \caption{Overview the pipeline of \textsc{FDINet}. 
  In the first step, we select $K$ anchor samples for each class $c$ (airplane in figure). 
  In the next step, we measure feature distortion to obtain FDI vector for each inspected sample.
  Finally, the extracted FDI vector is used to create a binary extraction attack detector and a colluding adversaries detector.}
  \label{fig:ProFedi_pipeline}
\end{figure*}

\subsection{Attack Objective and Assumptions}
\label{sec:attack_description}
\textbf{Adversary’s objective.}
In real-world scenarios, the adversary is typically restricted from accessing the inner operations of cloud-based models, including private training data, model architecture, and model parameters.
However, the adversary can still engage with the black-box model through the submission of queries and the retrieval of prediction probabilities via the publicly accessible API interface.
Without loss of generality, the adversary randomly selects samples from an unlabeled query set, submits these samples to the MLaaS platform to obtain predictions, and thereby acquires a labeled substitute set. Finally, the adversary uses this labeled substitute set to train their own substitute model.
Therefore, the objective of adversary can be summarized as:
\begin{equation}
    F_{S} = \mathop{max}\limits_{F_{S}} \{P_{(x\in\mathcal{D}_{test})}  \mathbbm{1}\left[{F}_{V}(x) = {F}_{S}(x)\right]\}
\end{equation}

\textbf{Adversary’s query set.}
We consider three types of adversaries as mentioned in Section~\ref{sec:related_atk} 
    (i.e., $\mathcal{A}_{sur}$, $\mathcal{A}_{adv}$, and $\mathcal{A}_{syn}$).
For $\mathcal{A}_{sur}$, the incoming queries come from PD and NPD natural images (e.g., Knockoff~\cite{orekondy2019knockoff}, ActiveThief~\cite{pal2020activethief}).
For $\mathcal{A}_{adv}$, the adversary generates adversarial examples from basic data with the aim of approximating the decision boundary of the target model (e.g., JBA~\cite{papernot2017practical}, T-RND~\cite{juuti2019prada}).
For $\mathcal{A}_{syn}$, the malicious queries are synthetic data from a generator (e.g., Data-free Substitute Training (DaST)~\cite{zhou2020dast}, 
Data-free Model Extraction (DFME)~\cite{truong2020data}).

\textbf{Colluding adversaries.}
Within the context of MLaaS, the basic defense strategy involves rejecting queries from suspected attackers. 
In this context, adversaries may employ multiple clients $N (N>1)$  to enhance the stealthiness of their attacks and bypass request limitations.
These colluding clients, under the control of a central adversary, collaborate to carry out model extraction attacks using similar query selection strategies, all working towards a common objective.
In this paper, we refer to these clients as colluding adversaries.

\textbf{Adaptive adversary.}
In the context of model extraction, we must consider the presence of an adaptive adversary who possesses knowledge of the defense methods employed.
This adversary can modify their query submission strategy to enhance the stealthiness of the extraction process.
In this paper, we focus on two types of adaptive adversaries:
(1) \textit{Dummy Query}: 
  This adaptive method, proposed by PRADA~\cite{juuti2019prada}, 
  involves generating dummy queries that do not contribute to the model extraction process.
  These queries are designed to maintain a normal distribution within a sequence of historical queries, thereby evading detection.
(2) \textit{Feature Correction}:
  To evade detection, the adversary employs an auxiliary encoder, denoted as $\hat{F}$, which is a pre-trained encoder drawn from model zoo. 
  This auxiliary encoder is used to correct the query’s feature maps before submitting them to the MLaaS system.
  We will discuss the detail of \textit{Feature Correction} in Section~\ref{sec:adaptive}.

\subsection{Defense Objective and Assumptions}
\label{sec:def_assumption}
Protection of user data is of utmost importance in the MLaaS platform. 
It is imperative for a secure MLaaS system to prioritize the confidentiality of user models.

\textbf{Defense objective.}
The defender acts as a crucial intermediary between the CSP and clients, 
  with the main goal of detecting and preventing extraction actions.
It aims to create a powerful extraction attack detector $\mathcal{C}$
  that can distinguish between benign and malicious queries.
As a result, the goal of defender is:
\begin{equation}
\resizebox{0.9\hsize}{!}{
$\mathop{max}\limits_{\mathcal{C}} {
    {P}_{(x \in \mathcal{D}_{adv})} \mathbbm{1}\left[\mathcal{C}({F}_{V};x)=1 \right] + 
    {P}_{(x \in \mathcal{D}_{test})} \mathbbm{1}\left[\mathcal{C}({F}_{V};x)=0 \right]},$
}
\end{equation}
where $\mathcal{D}_{adv}$ is query set of adversary, $\mathcal{C}$ is the extraction attack detector.
Additionally, the defender aims to evaluate the performance of detector using the following metrics: 
(1) \textit{effectiveness}, 
    the detector is expected to effectively identify various types of extraction attacks.
(2) \textit{efficiency}, 
    for the low-latency MLaaS platform, 
    the defense algorithm should immediately raise extraction alarms using only a few queries.
(3) \textit{robustness}, 
    the defender has the ability to resist stealthy attacks, 
    such as distributed attacks, as well as adaptive attacks.

\textbf{Defending assumptions.}
We consider the attack-agnostic scenario, 
    where the defender has no prior knowledge of sample selection strategies.
Besides, the attack model architecture, training mechanism, and relationship between clients are unknown to the defender.
We suppose that the defender knows developer's private training data $\mathcal{D}_{train}$
    and has access to the feature maps of victim model $F_{V}$.
The \textsc{ProFedi} is generic and flexible 
  since it makes no assumptions about the adversary, the DNN model, and the developer’s private training data.
This results in its defense capability of identifying extraction attacks in all scenarios,
  including $\mathcal{A}_{sur}$, $\mathcal{A}_{adv}$, and $\mathcal{A}_{syn}$.

\section{Methodology}
As shown in Figure~\ref{fig:ProFedi_pipeline},
    the detection process of \textsc{FDINet} includes the following phases: 
    (1) selecting anchor samples, 
    (2) measuring feature distortion, 
    (3) training extraction attack detector.
In the first phase, we select high prediction confidence data from the training set as anchor samples.
Subsequently, the feature space distances between each inspected sample and the anchor samples are calculated to generate the FDI vector.
Finally, the extracted FDI vector is employed as an intrinsic feature to train the extraction attack detector.

\subsection{Selecting Anchor Samples}
The adversary selects and/or crafts samples to query the public API, 
    then uses the prediction results as labels to train a clone model.
Intending to extract more information from the cloud-based model,
    the adversary explores enormous input space to increase the diversity of queried samples.
Therefore, the feature distributions of sample submitted by an adversary deviate from benign training data.
This phenomenon motivates us to design an effective metric
    to measure the feature distributions deviation between received query and training data.

The first step toward measuring feature space deviation is to select anchor samples.
To ensure the selected representative samples lie in the center, 
    we select $K$ high-confidence data from training set as anchor samples for each class.
It should be noted that these anchor samples lie at the center of each class, 
  encapsulating the statistical features of the benign query set.
Formally, for each class $c$, 
    the selected samples are denoted as ${\{(x_{c,j}, c)\}}_{j=1}^{K} \in \mathcal{D}_{anc}$.
Afterwards, we use the selected anchor sample $(x_{c,j}, c)$ to 
    extract feature maps ${F}_{V}^{l}\left(x_{c,j}\right)$ for layer $l$.

\subsection{Measuring Feature Distortion}
The observed deviation in feature distributions between an inspected sample and the anchor samples is referred to as the feature distortion of the inspected sample.
To quantitatively measure this feature distortion, 
  we introduce a novel metric called the Feature Distortion Index (FDI).
The FDI serves as a quantitative measure to assess the extent of feature distortion in the inspected sample compared to the anchor samples.
Formally, the FDI is defined as follows:
\begin{definition}[\textit{Feature Distortion Index}]
  \label{def:fdi}
  Given a victim model $F_{V}$, an anchor set $\mathcal{D}_{anc}$, 
    the feature distortion index for an inspected sample $(x, c)$ is defined as:
    \begin{equation}
      \mathcal{I}^{l}_{j} = d\left(F_{V}^{l}(x) - F_{V}^{l}(x_{c,j})\right) 
        \quad \text{s.t.} \, {x_{c,j} \in \mathcal{D}_{anc}},
    \end{equation}
  where $F_{V}^{l}(x)$ denotes the output feature of $F_{V}$ for layer $l$,
    $d$ indicates the $l2$-norm in our paper, 
    and $c$ is the label of $x$ predicted by ${F}_{V}$.
\end{definition}
Given a victim model $F_{V}$, we extract a total of $L$ layer feature maps (i.e., $l\in\{1,...,L\}$).
We then concatenate all $\mathcal{I}^{l}$ to obtain a $(K \times L)$-dimension FDI vector.
For example, for VGG19 of task CIFAR10, 
    we select $K=100$ anchor samples and extract $L=5$ layer feature maps
    to obtain a $(5 \times 100)$-dimension FDI vector $\mathcal{I}\left(x, c; \mathcal{D}_{anc}, F_{V}\right)$.

\subsection{Training Extraction Attack Detector}
\label{sec:method_detector}
In this section, we employ the extracted FDI vector as the inputs to 
  \textbf{(1)} construct a binary classifier to detect extraction queries,
  \textbf{(2)} perform hypothesis tests to identify colluding adversaries.

\begin{table}[!h]
  \centering
  \caption{Architecture of binary classifier $\mathcal{C}$}
  \begin{tabular}{l|c|c}
  \specialrule{1pt}{2pt}{2pt}
  \textbf{Index} & \textbf{Layer Type} & \textbf{Weight} \\
  \specialrule{1pt}{0pt}{0pt}
    1 & Linear & $[K \times L, 256]$ \\
    2 & Linear & $[256, 32]$ \\
    3 & Linear & $[32, 2]$ \\
    4 & Softmax & - \\
  \specialrule{1pt}{2pt}{2pt}
  \end{tabular}
  \label{tab:binary_classifier}
\end{table}

\subsubsection{Detecting Extraction Attacks}
Table~\ref{tab:binary_classifier} illustrates the architecture of $\mathcal{C}$.
The binary classifier, takes in a $(K \times L)$-dimension FDI vector as input and produces a $2$-dimensional probability vector as output.

\textbf{Training.} In order to train $\mathcal{C}$, 
  we adopt the training strategy commonly used in Out-of-Distribution detection.
Specifically, we gather a positive dataset and a negative dataset, 
  which are utilized during the training process.
In this paper, we utilize the $\mathcal{D}_{train}$ as the negative dataset 
  and collect an auxiliary dataset $\mathcal{D}_{aux}$ as the positive dataset.
The selection of $\mathcal{D}_{aux}$ will be discussed in Section~\ref{sec:setting_def}.

\textbf{Evaluation.}
Identifying malicious clients based on a single query can be challenging due to its low information entropy.
To overcome this limitation, we introduce a majority voting algorithm that utilizes a batch size of $bs$ queries to detect malicious clients.
In this approach, for each submitted query with a batch size of $bs$ samples, 
  we calculate the average confidence score of the predictions 
  and utilize it as an indicator to determine the maliciousness of the client.
\begin{equation}
    ac = \frac{1}{bs} \times \sum_{i=1}^{bs}{\mathop{\arg\max} \mathcal{C}(\mathcal{I}\left(x, c; \mathcal{D}_{anc}, F_{V}\right))},
\label{eqn:score}
\end{equation}
where $bs$ is the length of sequence and $ac \in [0, 1]$ is a confidence score.

Finally, we adopt a threshold $\tau_{1}$ to determine whether the queries come from malicious or benign clients.
Note that if confidence score $ac$ high than threshold $\tau_{1}$, the query is predicted as extraction attack.
The training and prediction of $\mathcal{C}$ are described in Algorithm~\ref{alg:binary_cls}.

\begin{algorithm}[!t]
  \caption{Training and Prediction of $\mathcal{C}$}
  \label{alg:binary_cls}
  \KwIn{Victim model ${F}_{V}$, 
      negative set $\mathcal{D}_{train}$, 
      positive set $\mathcal{D}_{aux}$, anchor set $\mathcal{D}_{anc}$,
      received query $x_{i} (i \leq bs)$}
  \KwOut{Majority voting score $ac$}
  \textsc{Function} Training: \\
  \For{$(x, y) \gets \mathcal{D}_{train} \cup \mathcal{D}_{aux}$}{
      $c = {F}_{V}(x)$ \\
      $\mathcal{I} = [\mathcal{I}_{1}^{1},...,\mathcal{I}_{K}^{1},...,\mathcal{I}_{1}^{L},...,\mathcal{I}_{K}^{L}]$ \\
      $\mathcal{C} = \mathop{\arg\min}\limits_{\mathcal{C}} \mathcal{L}_{\text{cross-entropy}} (\mathcal{I}, y)$ // training step \\
  }
  \textsc{Function} Prediction: \\
  $score = 0$ \\
  \For{$i \gets bs$}{
    $c_{i} = {F}_{V}(x_{i})$ \\
    $\mathcal{I} = [\mathcal{I}_{1}^{1},...,\mathcal{I}_{K}^{1},...,\mathcal{I}_{1}^{L},...,\mathcal{I}_{K}^{L}]$ \\
    $score += \mathop{\arg\max} \mathcal{C}(\mathcal{I})$
  }
  $ac = score/bs$ \\
  \Return{$ac$}
\end{algorithm}

\subsubsection{Identifying Colluding Adversaries}
In this section, we first define distributed model extraction attacks and introduce our method to identify colluding adversaries.

\textbf{Distributed extraction attack.}
Given an MLaaS platform with $M=\{1, 2,...,M\}$ clients, a central adversary controls a set of $N (2 \leq N \leq M)$ clients.
The central adversary adopts sample selection strategies (e.g., $\mathcal{A}_{sur}$, $\mathcal{A}_{adv}$, and $\mathcal{A}_{syn}$) 
  to build a query set and distributes them to each client.
The distributed attack is stealthy since each controlled agent only has a small overhead that is easy to evade request limitations.

\textbf{Colluding adversaries detection.}
We observe a significant level of FDI similarity among queries that are generated using the same model extraction attack.
This key observation motives us to detect colluding adversaries using FDI similarity.

In order to determine if two examined clients, $u$ and $v$, are colluding adversaries, 
  our approach involves collecting a set of $n \times bs$ historical queries from each client.
Subsequently, we extract their FDI vectors and perform two-sample hypothesis tests for further analysis as follows:
\begin{proposition}[\textit{Two-sample Hypothesis Tests}]
  \label{prop:hypothesis}
  Given two inspected clients $u$ and $v$, and their $n \times bs$ FDI vectors $\mathcal{I}_{u}$ and $\mathcal{I}_{v}$,
  the null hypothesis can be expressed as: $\mathcal{H}_{0}: \mu_{u} = \mu_{v}$,
  while the alternative hypothesis is expressed as $\mathcal{H}_{a}: \mu_{u} \neq \mu_{v}$.
  Though calculating the test statistic $\mathbf{t}=\left(\bar{x}_1-\bar{x}_2\right) / {s}_{p}\left(\sqrt{\frac{1}{|\mathcal{I}_{u}|} + \frac{1}{|\mathcal{I}_{v}|}}\right)$, 
  we can obtain its p-value, where $\bar{x}_1$ and $\bar{x}_2$ are means of samples, ${s}_{p}$ indicates variance.
\end{proposition}
Finally, we select a threshold $\tau_{2}$ chosen for statistical significance (usually $\tau_{2} = 0.05$) for testing.
If the calculated p-value is below $\tau_{2}$, 
  then the null hypothesis $\mathcal{H}_{0}$ is rejected in favor of the alternative hypothesis $\mathcal{H}_{a}$.
In this case, it indicates that clients $u$ and $v$ are not colluding adversaries.

\subsection{Adaptive Attacks}
\label{sec:adaptive}
An adaptive adversary who knows \textsc{FDINet} 
    may potentially modify attack strategies to evade our detection.
In this section, we assume that 
    the adaptive adversary has a mini-batch of substitute anchor samples
    and an auxiliary encoder $\hat{F}$ drawn from the model zoo.
We proposed \textit{Feature Correction (FeatC)}, an adaptive attack that
    utilizes an auxiliary encoder to make the query similar to anchors samples in feature maps.
Formally, the adaptive adversary locally perturbates $x$ using loss function $\mathcal{L}$:
\begin{align}
\mathcal{L}(x; \hat{F}, \hat{x}) &= {||\hat{F}^{l}(x+\delta) - \hat{F}^{l}(\hat{x})||}_{2}^{2} \\ 
  &\text{s.t.} \quad F_{V}(x)=F_{V}(\hat{x}), \delta < \epsilon, \notag
\end{align}
where $\epsilon$ refers the attack bound, $\hat{F}$ is a pre-trained feature extractor drawn from model zoo 
    and $\hat{x}$ denotes auxiliary anchor sample from victim training data $\mathcal{D}_{train}$.
Through generating feature-corrected queries, 
    the adaptive adversary intends to bypass our detection.
Since \textit{FeatC} exploits the full knowledge of our defense mechanism,
    we believe \textit{FeatC} is a strong adaptive attack against \textsc{FDINet}.

\section{Experiments}
In this section, we conduct extensive experiments to validate the performance of \textsc{FDINet} 
  against six advanced model extraction attacks, covering four different deep learning systems.
We begin by introducing the experimental setup in Section~\ref{sec:exp_setup}. 
Subsequently, we evaluate the performance of \textsc{FDINet} against extraction attacks in Section~\ref{sec:exp1} and distributed attacks in Section~\ref{sec:exp2}. 
Additionally, we conduct ablation studies in Section~\ref{sec:exp3} and explore the adaptive attacks in Section~\ref{sec:exp4}.
All experiments are performed on an Ubuntu 20.04 system equipped with a 96-core Intel CPU and four Nvidia GeForce GTX 3090 GPU cards. 
The machine learning algorithm is implemented using PyTorch v1.10.

\subsection{Experimental Setup}
\label{sec:exp_setup}
\subsubsection{Datasets and Victim Models}
\begin{table}[!h]
  \centering
  \caption{Datasets and victim models.}
  \resizebox{\linewidth}{!}{
    \begin{tabular}{l|c|c|c}
        \specialrule{1pt}{2pt}{2pt}
        \textbf{Dataset}  & \textbf{Model} & \textbf{Acc.(\%)} & \textbf{Scenario} \\
        \specialrule{1pt}{0pt}{0pt}
        CIFAR10     & VGG19 & 87.73 & General Visual Recognition \\ 
        \hline
        GTSRB       & MobileNetV2 & 90.99 & Traffic Sign Recognition \\ 
        \hline
        CelebA      & DenseNet121 & 93.49 & Face Recognition \\ 
        \hline
        Skin Cancer & ResNet50 & 98.52 & Disease Diagnosis\\ 
        \specialrule{1pt}{2pt}{2pt}
    \end{tabular}
  }
  \label{tab:exp_setting}
\end{table}
Our method is assessed on four benchmark datasets: 
  CIFAR10~\cite{krizhevsky2009learning}, GTSRB~\cite{stallkamp2012man}, 
  CelebA~\cite{liu2018large}~\footnote{We adopt gender attributes in our experiment.}, and Skin Cancer~\cite{codella2019skin}.
These datasets cover four distinct deep learning systems that are commonly employed in security-critical domains: 
  general visual recognition, traffic sign recognition, face recognition, and disease diagnosis.
To conduct our experiments, we utilize four different convolutional neural networks: VGG19, MobileNetV2, DenseNet121, and ResNet50.
To accommodate the input size of $32\times32$, 
  we adjusted the filter size of the first convolutional layer in the original architecture by downsampling. 
Table~\ref{tab:exp_setting} presents a summary of the datasets and victim models utilized in our experiments.

\subsubsection{Setting of Attack Methods}
\begin{table}[!h]
  \centering
  \caption{Surrogate model’s Top-3 accuracy(\%) of various model extraction attacks for 
  $\mathcal{A}_{sur}$, $\mathcal{A}_{adv}$ and $\mathcal{A}_{syn}$ scenarios.}
  \resizebox{\linewidth}{!}{
    \begin{tabular}{l|cc|cc|cc}
        \specialrule{1pt}{2pt}{2pt}
        \multirow{2}{*}{\textbf{Dataset}} &
        \multicolumn{2}{c|}{\textbf{$\mathcal{A}_{adv}$}} & 
        \multicolumn{2}{c|}{\textbf{$\mathcal{A}_{sur}$}} &
        \multicolumn{2}{c}{\textbf{$\mathcal{A}_{syn}$}} \\
        \cline{2-7}
        & \textbf{JBA} & \textbf{T-RND} & \textbf{Knockoff} & \textbf{ActiveThief} & \textbf{DFME} & \textbf{DaST} \\
        \specialrule{1pt}{0pt}{0pt}
        CIFAR10 & 54.87 & 62.61 & 97.23 & 95.23 & 96.98 & 35.18\\
        \hline
        GTSRB & 26.79 & 21.65 & 34.26 & 33.99 &  92.73 & 40.09 \\
        \hline
        CelebA & 76.55 & 76.94 & 99.85 & 82.36 &  92.74 & 42.84 \\
        \hline
        Skin Cancer & 74.44 & 69.50 & 94.89 & 99.48 &  82.56 & 61.35 \\
        \specialrule{1pt}{2pt}{2pt}
    \end{tabular}
  }
  \label{tab:atk_setting}
\end{table}
Six advanced model extraction attacks are considered in our experiments, 
  covering three adversarial scenarios, i.e., $\mathcal{A}_{adv}$, $\mathcal{A}_{sur}$ and $\mathcal{A}_{syn}$ 
  (as mentioned in Section~\ref{sec:related_atk}).
For $\mathcal{A}_{adv}$ scenario,
  we assume the adversary has a mini-batch of victim's training data 
  and employs Jacobian-based Augmentation (JBA) ~\cite{papernot2017practical} 
  \underline{T}argeted \underline{Ran}domly Chosen Diraction (T-RND)~\cite{juuti2019prada} to create extraction query.
For $\mathcal{A}_{sur}$ scenario,
  we follow the experiment setting of Knockoff~\cite{orekondy2019knockoff},
  which assumes the adversary selects queries from a surrogate dataset.
Specifically, we adopt CINIC-10~\cite{darlow2018cinic}, 
  TSRD~\footnote{http://www.nlpr.ia.ac.cn/pal/trafficdata/recognition.html}, 
  LFW~\cite{huang2008labeled}, and BCN20000~\cite{combalia2019bcn20000} as
    Knockoff and ActiveThief~\cite{pal2020activethief}’s 
    surrogate dataset for the tasks of CIFAR10, GTSRB, CelebA and Skin Cancer, respectively.
For $\mathcal{A}_{syn}$ scenario,
  we follow DFME~\cite{truong2020data} and DaST~\cite{zhou2020dast}'s experiment setting 
    that employs a generative model to craft surrogate data as query set.
Table~\ref{tab:atk_setting} provides a summary of the surrogate model's Top-3 accuracy for each attack.
Note that those six model extraction attacks compass a wide range of cutting-edge techniques, 
  and their queries cover problem domain data, non-problem domain data, adversarial examples, and synthetic data.

\begin{table*}[!h]
    \small
    \centering
    \caption{The DAcc. and FPR of \textsc{FDINet}, PRADA and SEAT 
    to detect against the-state-of-art extraction attacks.}
    \resizebox{0.85\linewidth}{!}{
        \begin{tabular}{cccc|cc|cc|cc|cc|cc|cc}
            \specialrule{1pt}{2pt}{2pt}
            \multirow{2}{*}{\textbf{Dataset}} & \multirow{2}{*}{\textbf{Strategy}} 
            & \multirow{2}{*}{$bs$} & \multirow{2}{*}{$\tau_{1}$}
            & \multicolumn{2}{c|}{\textbf{JBA}} & \multicolumn{2}{c|}{\textbf{T-RND}} 
            & \multicolumn{2}{c|}{\textbf{Knockoff}} & \multicolumn{2}{c|}{\textbf{ActiveThief}} 
            & \multicolumn{2}{c|}{\textbf{DFME}} & \multicolumn{2}{c}{\textbf{DaST}} \\
            \cline{5-16}
            & & & & DAcc. & FPR & DAcc. & FPR & DAcc. & FPR & DAcc. & FPR & DAcc. & FPR & DAcc. & FPR \\
            \specialrule{1pt}{0pt}{0pt}
          
            \multirow{7}{*}{\rotatebox{90}{\textbf{CIFAR10}}}
            & \multirow{2}{*}{SEAT}
            &   50  & 0.87
                & 9.30 & 0.20 & 11.40 & 0.30 
                & 5.80 & 0.0 & 5.30 & 0.0
                & 12.20 & 0.10 & 13.90 & 0.0\\
            & & 500 & 0.81
                & 91.00 & 0.0 & 90.00 & 0.0 
                & 55.00 & 0.0 & 45.00 & 0.0 
                & 75.00 & 0.0 & 77.00 & 0.0\\ 
            \cline{3-16}
            & \multirow{2}{*}{\shortstack{PRADA}}
            &   50  & 0.99 
              & 0.0 & 0.0 & 0.0 & 0.0 
              & 0.0 & 0.0 & 0.0 & 0.0 
              & 0.0 & 0.0 & 0.0 & 0.0 \\
            & & 500 & 0.97 
              & \textbf{100.0} & 36.00 & \textbf{100.0} & 36.00
              & 28.00 & 36.00 & 39.00 & 36.00
              & 30.00 & 36.00 & 19.00 & 0.0 \\
            \cline{3-16}
            & \multirow{2}{*}{\shortstack{\textsc{FDINet}}}
            & 50 & 0.47 
                & 89.90 & 4.10 & 93.10  & 2.60
                & 80.26 & 4.10 & 82.00  & 2.90
                & \textbf{100.0} & 1.70 & 98.03 & 1.80 \\
            & & 500 & 0.48
                & 91.00 & 0.0 & 93.00 & 0.0
                & 90.00 & 0.0 & 92.00 & 0.0
                & \textbf{100.0} & 0.0 & \textbf{100.0} & 0.0\\
            \specialrule{1pt}{1pt}{1pt}

            \multirow{7}{*}{\rotatebox{90}{\textbf{GTSRB}}}
            & \multirow{2}{*}{SEAT}
            &   50  & 0.90
                & 4.20 & 4.40 & 3.80 & 4.40 
                & 8.20 & 4.20 & 8.60 & 4.20 
                & 34.20 & 4.40 & 43.20 & 4.40\\
            & & 500 & 0.88
                & 74.00 & 4.00 & 77.00 & 4.00 
                & 65.00 & 4.00 & 55.00 & 4.00 
                & 89.00 & 4.00 & 85.00 & 4.00\\ 
            \cline{3-16}
            & \multirow{2}{*}{\shortstack{PRADA}}
            &   50  & 0.94
                & 0.0 & 0.0 & 0.0 & 0.0 
                & 0.0 & 0.0 & 0.0 & 0.0 
                & 0.0 & 0.0 & 0.0 & 0.0 \\
            & & 500 & 0.93
                & 97.00 & 9.00 & 97.00 & 9.00 
                & 0.00 & 9.00 & 1.00 & 9.00 
                & 1.00 & 9.00 & 16.00 & 9.00 \\
            \cline{3-16}
            & \multirow{2}{*}{\shortstack{\textsc{FDINet}}}
            &   50  & 0.77
                & 90.30 & 0.0 & 90.40 & 0.0 
                & \textbf{100.0} & 0.0 & 95.60 & 0.0 
                & \textbf{100.0} & 0.0 & \textbf{100.0} & 0.0\\
            & & 500 & 0.78
                & 88.00 & 0.0 & 87.00 & 0.0 
                & \textbf{100.0} & 0.0 & 95.00 & 0.0 
                & \textbf{100.0} & 0.0 & \textbf{100.0} & 0.0\\ 
            \specialrule{1pt}{1pt}{1pt}

            \multirow{7}{*}{\rotatebox{90}{\textbf{CelebA}}}
            & \multirow{2}{*}{SEAT}
            &   50  & 0.90
                & 5.50 & 0.0 & 8.40 & 0.0 
                & 0.0 & 0.0 & 0.0 & 0.0 
                & 14.10 & 0.0 & 19.30 & 0.0\\
            & & 500 & 0.82
                & 89.00 & 0.0 & 79.00 & 0.0
                & 0.0 & 0.0 & 0.0 & 0.0 
                & 79.00 & 0.0 & 90.00 & 0.0\\ 
            \cline{3-16}
            & \multirow{2}{*}{\shortstack{PRADA}}
            &   50  & 0.96
                & 0.0 & 0.0 & 0.0 & 0.0 
                & 0.0 & 0.0 & 0.0 & 0.0 
                & 0.0 & 0.0 & 0.0 & 0.0 \\
            & & 500 & 0.95
                & \textbf{100.0} & 6.00 & \textbf{100.0} & 6.00 
                & 3.00 & 6.00 & 1.00 & 6.00 
                & 5.00 & 6.00 & 9.00 & 6.00\\ 
            \cline{3-16}
            & \multirow{2}{*}{\shortstack{\textsc{FDINet}}}
            &   50  & 0.33
                & 96.61 & 0.30 & 97.10 & 0.10 
                & 83.93 & 0.30 & 97.10 & 0.0 
                & \textbf{100.0} & 0.10 & \textbf{100.0} & 0.20\\ 
            & & 500 & 0.36
                & 95.00 & 0.0 & 96.00 & 0.0 
                & 93.00 & 0.0 & 93.00 & 0.0 
                & \textbf{100.0} & 0.0 & \textbf{100.0} & 0.0\\
            \specialrule{1pt}{1pt}{1pt}

            \multirow{7}{*}{\rotatebox{90}{\textbf{Skin Cancer}}}
            & \multirow{2}{*}{SEAT}
            &   50  & 0.90
                & 10.20 & 0.0 & 8.90 & 0.0 
                & 5.30 & 0.0 & 6.00 & 0.0 
                & 4.50 & 0.0 & 8.40 & 0.0\\
            & & 500 & 0.82
                & 69.00 & 2.00 & 73.00 & 2.00 
                & 62.00 & 2.00 & 63.00 & 2.00 
                & 81.00 & 2.00 & 87.00 & 2.00 \\ 
            \cline{3-16}
            & \multirow{2}{*}{\shortstack{PRADA}}
            &   50  & 0.93
                & 0.0 & 0.0 & 0.0 & 0.0 
                & 0.0 & 0.0 & 0.0 & 0.0 
                & 0.0 & 0.0 & 0.0 & 0.0 \\
            & & 500 & 0.91
                & 98.00 & 1.00 & 98.00 & 1.00 
                & 28.00 & 1.00 & 26.00 & 1.00 
                & 23.00 & 1.00 & 34.00 & 1.00\\ 
            \cline{3-16}
            & \multirow{2}{*}{\shortstack{\textsc{FDINet}}}
            &   50  & 0.51
                & 92.80 & 0.0 & 93.20 & 0.0 
                & 88.70 & 0.0 & 89.90 & 0.0 
                & 99.80 & 0.0 & 97.40 & 0.0\\
            & & 500 & 0.48
                & 94.00 & 0.0 & 94.00 & 0.0 
                & \textbf{100.0} & 0.0 & 80.00 & 0.0 
                & \textbf{100.0} & 0.0 & \textbf{100.0} & 0.0 \\ 
            \specialrule{1pt}{2pt}{2pt}
        \end{tabular}}
    \label{tab:exp11_detection}
\end{table*}

\subsubsection{Setting of Defense Methods}
\label{sec:setting_def}
In the main paper, we conduct a performance comparison between 
  \textsc{FDINet}, PRADA~\cite{juuti2019prada}, SEAT~\cite{zhang2021seat} and Extraction Monitor~\cite{kesarwani2018model}.
To ensure consistency, we utilize the official implementation of PRADA 
  and make adjustments to the hyperparameter $\tau_{1}$ in order to achieve a low false positive rate (FPR) on the validation set.
Regarding SEAT, we employ the victim model as an encoder and perform fine-tuning for 20 epochs 
  using the Stochastic Gradient Descent (SGD) optimizer with a learning rate of 0.001.
Following SEAT's methodology, we select a threshold that yields a low FPR on the validation set.
For the Extraction Monitor, we adopt the same architecture as the victim model and treat it as a proxy model, as suggested in the original paper. 
To train the proxy model, we utilize SGD with a learning rate of 0.005 for 2 iterations per batch of submitted queries.

When considering \textsc{FDINet}, we establish the number of anchor samples ($K$) as 20 for GTSRB, CelebA, and Skin Cancer datasets, whereas for CIFAR10, $K$ is set to 100. 
We divide the neural network into multiple convolutional blocks and extract 5 layers (i.e., $L=5$) for all tasks.
As for the auxiliary dataset $\mathcal{D}_{aux}$, 
  we employ the testing set from CINIC-10, TSRD, VGGFace2, and BCN20000 
  as negative data for CIFAR10, GTSRB, CelebA, and Skin Cancer, respectively. 
It is important to note that the auxiliary datasets are based on realistic assumptions, 
  and the testing set does not overlap with the surrogate set used by the attacker.
To detect distributed attacks, we utilize two-sample hypothesis tests for the two inspected clients, denoted as $u$ and $v$.

\subsubsection{Evaluation Metrics}
In the experiments, we utilize five commonly employed metrics to assess the efficacy of our method: 
  Detection Accuracy (DAcc.), False Positive Rate (FPR), Extraction Status (ES), Colluding Detection Accuracy (CDAcc.), and p-value of hypothesis tests.
We will discuss the detail of each metric in the following experiments.

\subsubsection{Threshold Selection}
\label{sec:threshold}
Achieving optimal DAcc. and FPR in binary classification tasks relies heavily on selecting the right threshold. 
Inspired by previous research~\cite{juuti2019prada}, we introduce a data-driven threshold selection strategy. 
Initially, we utilize the validation set $\mathcal{D}_{val}$ to calculate the values of ${\mu}_{ac}$ and ${\sigma}_{ac}$, and subsequently apply the $3\sigma$ rule to set $\tau_{1}={\mu}_{ac} + 3 \times {\sigma}_{ac}$.
We will discuss the impacts of threshold in Section~\ref{sec:exp_threshold}.

\subsection{Detecting Extraction Attacks}
\label{sec:exp1}
In this experiment, we launch extraction attacks and generate $50,000$ samples as malicious client’s query set $\mathcal{D}_{adv}$.
We evaluate $\textsc{FDINet}$ using DAcc., FPR, and extraction status.
The DAcc. serves as a measure of accuracy for detecting malicious queries within the MLaaS system.
On the other hand, the FPR quantifies the rate at which negative samples are erroneously classified as positive by the binary detector.
It helps evaluate the system's performance in terms of misclassifying negative instances.
Additionally, the ES metric evaluates the fidelity between the proxy model and the victim model.

\subsubsection{Detection Accuracy and FPR}
\begin{figure}[!t]
  \centering
  \includegraphics[width=0.9\linewidth]{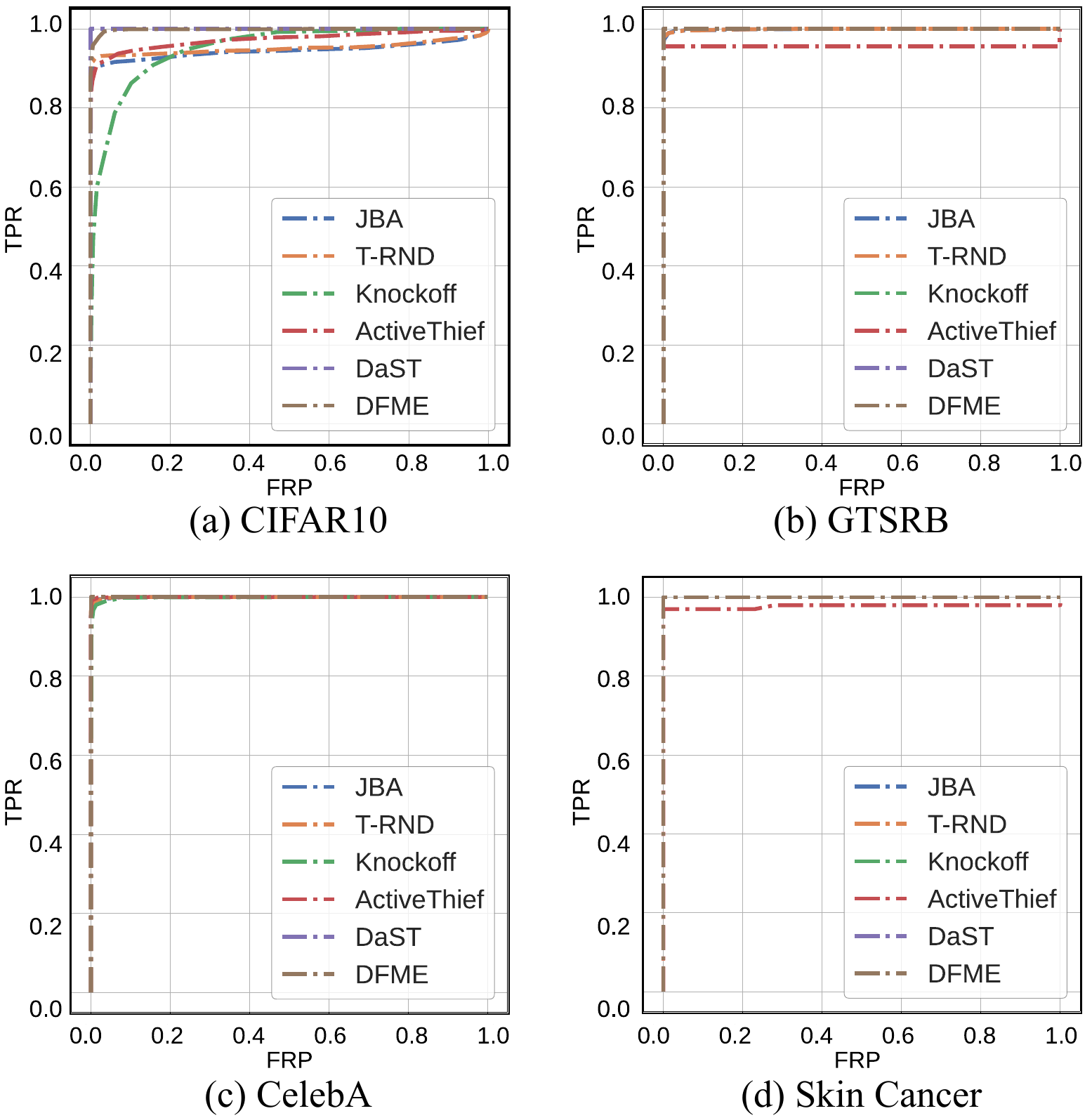}
  \caption{ROC curve of model extraction attacks detection.}
  \label{fig:exp12_roc}
\end{figure}

Table~\ref{tab:exp11_detection} presents a summary of the performance comparison between \textsc{FDINet}, PRADA, and SEAT against six model extraction attacks.
In our experiments, we examine two query batch sizes, $bs=50$ and $bs=500$.
Figure~\ref{fig:exp12_roc} illustrates the ROC curve of our method for detecting extraction queries.

In terms of performance, \textsc{FDINet} outperforms PRADA and SEAT with high DAcc. and low FPR. 
Specifically, \textsc{FDINet} achieves a DAcc. of $\textbf{100\%}$ and an FPR close to \textbf{0.0} for DFME and DaST with a batch size ($bs$) of 500. 
Furthermore, \textsc{FDINet} is capable of identifying malicious clients with just 50 queries.
On the other hand, both PRADA and SEAT fail to detect extraction attacks when $bs$ is set to 50.
It should be noted that \textsc{FDINet} achieves a lower DAcc. in CIFAR10 for Knockoff and ActiveThief.
This is because the surrogate dataset (CINIC-10) used by Knockoff and ActiveThief has some overlap with CIFAR10.
In the ablation study, we will explore the effects of threshold $\tau_{1}$ and batch size (bs) further, which can be found in Section~\ref{sec:exp3}.

\subsubsection{Extraction Status (ES)}
ES serves as another important metric for detecting extraction queries.
It is a metric introduced by EM~\cite{kesarwani2018model}, which employs information gain to quantify the level of model privacy leakage from the victim model.
EM utilizes a local proxy model to monitor the information gain of each client.
When the proxy model learns a surrogate model with high fidelity, extraction warnings are sent to the Cloud Service Provider (CSP).
Formally, the ES is defined as:
\begin{equation}
    es = \frac{100}{|\mathcal{D}_{test}|} \sum_{x \in \mathcal{D}_{test}} \mathbbm{1}\left[{F}_{V}(x) = {F}_{V^{'}}(x)\right].
\end{equation}
Since \textsc{FDINet} doesn't make use of proxy model, 
    we use average confidence (i.e., $\left(100 \times ac\right)\%$) as ES for comparison.

\begin{figure}[!t]
  \centering
  \includegraphics[width=0.9\linewidth]{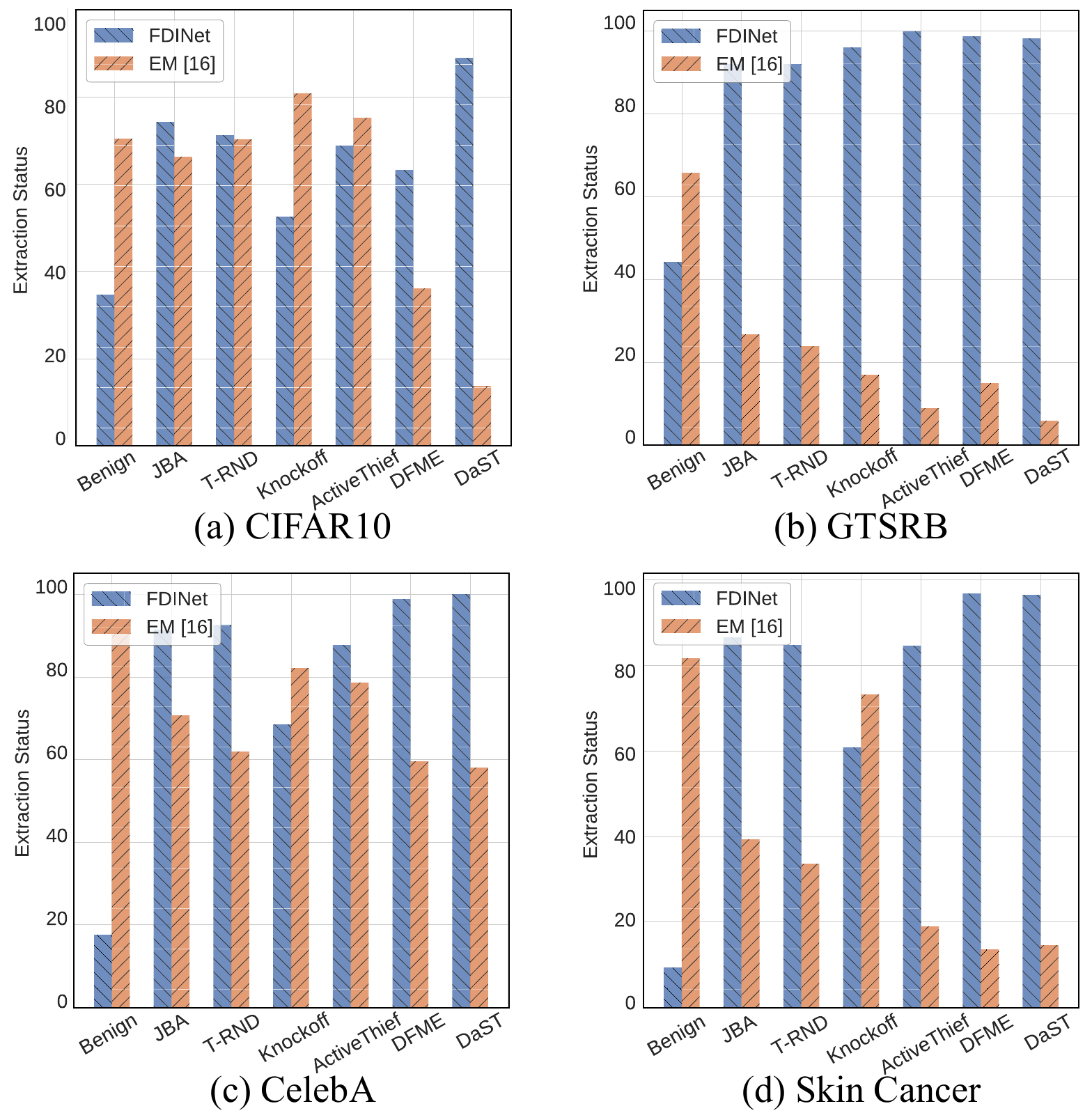}
  \caption{Results of average Extraction Status for benign and malicious clients (\textbf{lower is better for benign clients}).
  Extraction Status (ES) is a metric proposed by ~\cite{kesarwani2018model} that uses information gain to quantify model privacy leakage from the victim model.}
  \label{fig:es_curve}
\end{figure}
Figure~\ref{fig:es_curve} depicts the average ES reported by \textsc{FDINet} and EM for both benign and malicious clients.
The results demonstrate that \textsc{FDINet}'s ES is 
  \textbf{below 34.60\% for CIFAR10}, \textbf{44.20\% for GTSRB}, \textbf{17.40\% for CelebA}, and \textbf{9.30\% for Skin Cancer}.
Additionally, we achieve an average of \textbf{96.08\% and 84.94\% for GTSRB and Skin Cancer}, respectively.
On the other hand, \textsc{FDINet} effectively identifies high ES for malicious clients (i.e., JBA, T-RND, Knockoff, ActiveThief, DFME, and DaST).
In contrast, EM reports high ES for benign clients due to their significant information gain.
Furthermore, the ES of EM is very low for DFME and DaST since these synthetic data samples have low information gain.
It is important to note that lower ES is preferable for benign queries, while higher ES is better for malicious queries.
Therefore, while EM effectively detects certain extraction attacks, it may also produce considerable false alarms for benign queries.

\begin{figure*}[!t]
  \centering
  \includegraphics[width=0.97\linewidth]{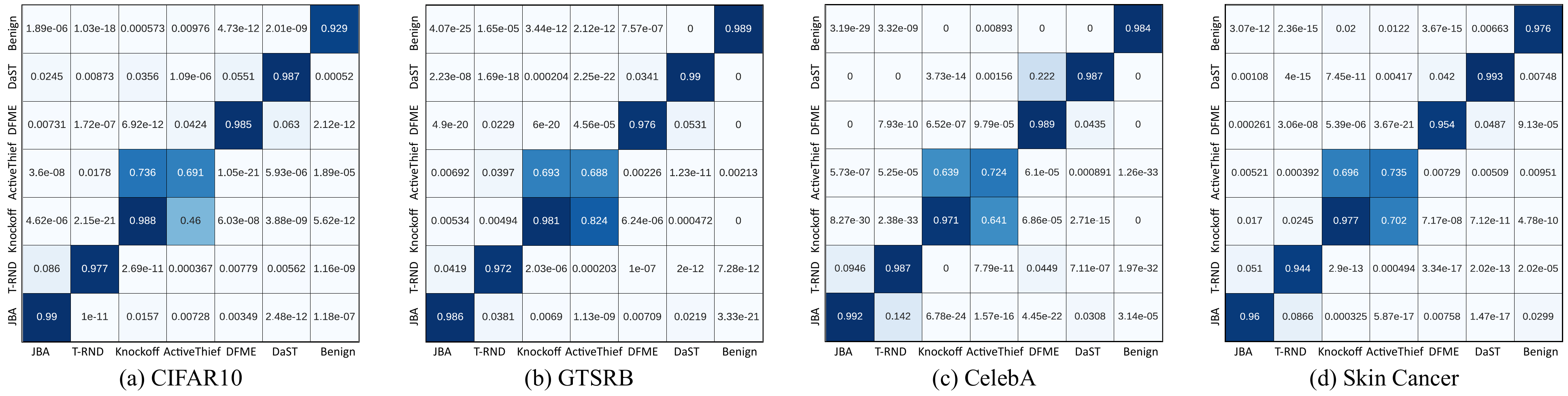}
  \caption{Illustration of the confusion matrix for average hypothesis tests' p-values over different clients.
  If the p-value is higher than 0.05, we accept $\mathcal{H}_{0}$, meaning clients $u$ and $v$ are colluding adversaries.}
  \label{fig:exp2.1_colluding_cm}
\end{figure*}

\subsubsection{Memory Costing and Detection Fine-grained}
Efficiency is crucial in the context of MLaaS, particularly when dealing with real-time APIs.
In the case of security-focused MLaaS, efficiency encompasses two key aspects: 
  resource consumption and detection fine-grained.
To assess our method’s efficiency, we compare it with state-of-the-art defense methods.

First, PRADA, relies on calculating the L2 distance between new and previous samples to identify malicious queries.
However, this approach necessitates significant memory storage.
Additionally, EM requires the maintenance of a local proxy model for each client, resulting in substantial computational overhead.
In contrast, our approach, \textsc{FDINet}, is lightweight and flexible. It doesn't depend on historical queries and doesn't make assumptions about the victim model.
We conducted an experiment using 50,000 testing queries.
The results demonstrate that \textsc{FDINet} achieves a throughput of 838.36 queries on the task of CIFAR10.
This showcases the efficiency of \textsc{FDINet} in processing queries promptly and effectively.

Furthermore, as shown in Table~\ref{tab:exp11_detection}, \textsc{FDINet} is efficient in identifying extraction queries using only 50 queries.
This highlights the efficiency of \textsc{FDINet} in swiftly and accurately detecting adversaries, thereby maximizing the protection of the victim model.

\begin{figure}[!h]
  \centering
  \includegraphics[width=0.9\linewidth]{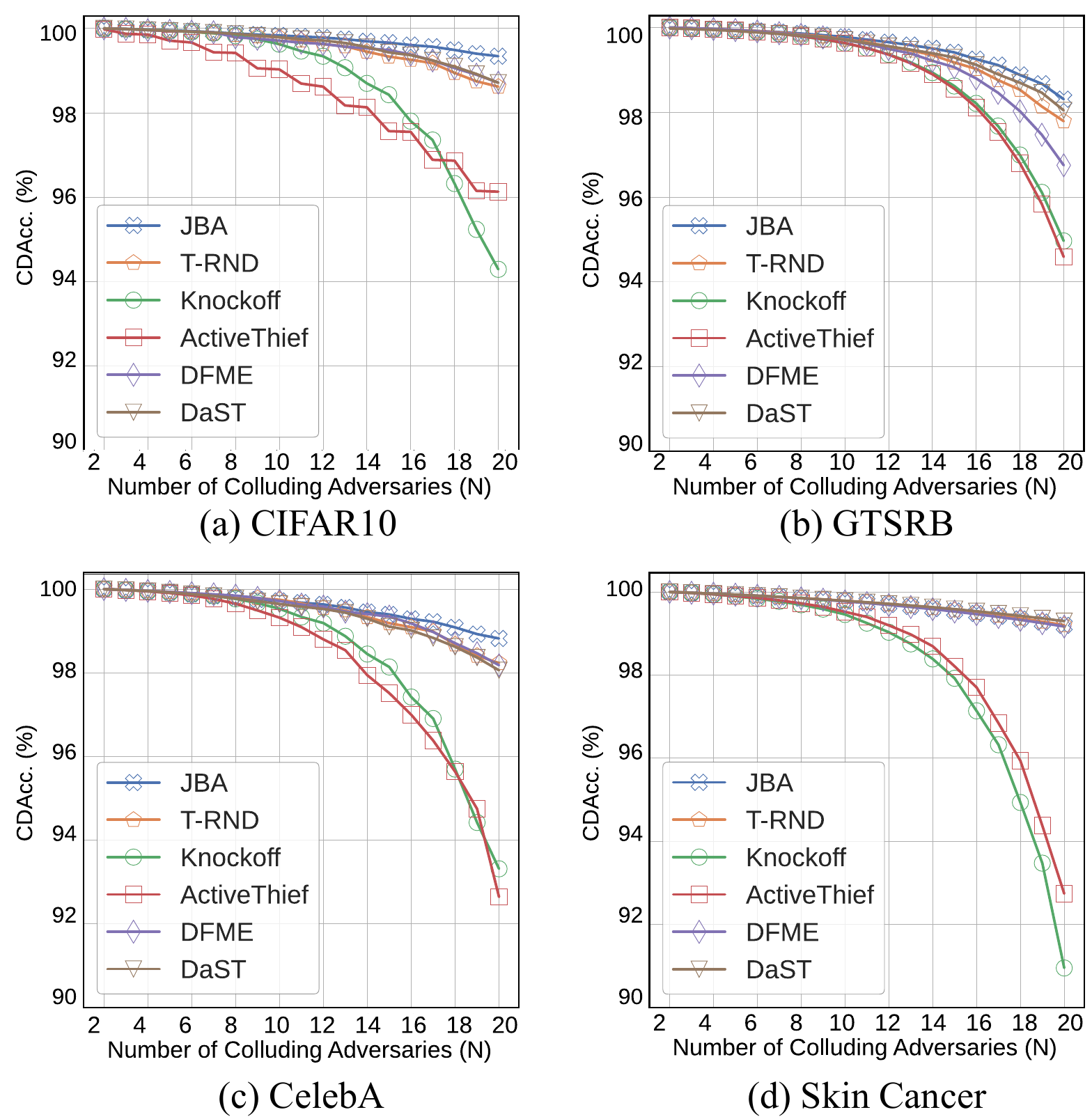}
  \caption{Performance of colluding adversaries detection for distributed attacks.
  We consider a 100 clients MLaaS platform.
  Among them, $2 \sim 20$ are colluding adversaries for each attack.}
  \label{fig:exp2.2_CDAcc_curve}
\end{figure}

\subsection{Detecting Distributed Extraction Attacks}
\label{sec:exp2}
In distributed extraction attack, the adversary distributes malicious queries to $N (N>1)$ clients.
The primary goal of \textsc{FDINet} is not to identify malicious queries, but rather to identify colluding adversaries.
To evaluate the performance of \textsc{FDINet}, we simulate an MLaaS system with $M=100$ clients, each submitting $50,000$ queries.
Among them, $2 \sim 20$ are colluding adversaries who jointly launch the same extraction attack.
In this experiment, we set $bs=500$ and $n=100$, with a total of $50,000$ samples.
For each pair of clients under inspection, denoted as $u$ and $v$, we extracted their FDI vectors.
Subsequently, we conducted two-sample hypothesis tests (as discuss in Proposition~\ref{prop:hypothesis}) to determine whether these clients were colluding adversaries.
This process allowed us to identify and expose potential collusive behavior among the clients in the MLaaS system.
The Colluding Detection Accuracy (CDAcc.) can be formulated as:
\begin{equation}
\resizebox{0.9\hsize}{!}{
    $\text{CDAcc.}=\frac{\sum_{u<v, u \in [1, N], v \in [1, N]} 
    \mathbbm{1}[\textsc{FDINet}(u, v) = J(u, v)] }{C_{N}^{2}} \times 100\%,$ 
}
\end{equation}
where $C$ is combination formula (i.e., $C_{n}^{m}=\frac{n!}{m!(n-m)!}$),
$J$ is the judgment function that returns $1$ if client $u$ and client $v$ are colluding adversaries.
In order to speed up the detection of \textsc{FDINet}, we first use binary detection to filter the benign clients,
then setup two-sample hypothesis tests.

\begin{table*}[!h]
    \small
    \centering
    \caption{The DAcc., FPR of \textsc{FDINet} to defend against \textit{FeatC}.}
    \resizebox{0.85\linewidth}{!}{
        \begin{tabular}{cc|cc|cc|cc|cc|cc|cc}
            \specialrule{1pt}{2pt}{2pt}
            \multirow{2}{*}{\textbf{Dataset}} & \multirow{2}{*}{$\hat{F}$} 
            & \multicolumn{2}{c|}{\textbf{JBA}} & \multicolumn{2}{c|}{\textbf{T-RND}} 
            & \multicolumn{2}{c|}{\textbf{Knockoff}} & \multicolumn{2}{c|}{\textbf{ActiveThief}} 
            & \multicolumn{2}{c|}{\textbf{DFME}} & \multicolumn{2}{c}{\textbf{DaST}} \\
            \cline{3-14}
            & & DAcc. & FPR & DAcc. & FPR & DAcc. & FPR & DAcc. & FPR & DAcc. & FPR & DAcc. & FPR \\
            \specialrule{1pt}{0pt}{0pt}
          
            \multirow{2}{*}{CIFAR10}
            & VGG11 
              & 90.30 & 6.10 & 93.10 & 5.70 
              & 51.30 & 4.70 & 64.00 & 6.10 
              & \textbf{100.0} & 5.70 & 98.90 & 4.70 \\
            & ResNet50 
              & 91.60 & 4.00 & 93.30 & 3.40 
              & 32.10 & 2.80 & 43.30 & 4.00 
              & \textbf{100.0} & 3.40 & 98.70 & 2.80\\
            \hline

            \multirow{2}{*}{GTSRB}
            & VGG11 
              & 86.60 & 0.0 & 84.50 & 0.0
              & 99.60 & 0.0 & 99.90 & 0.0 
              & 99.90 & 0.0 & \textbf{100.0} & 0.0 \\
            & ResNet50 
              & 86.60 & 0.0 & 84.50 & 0.0
              & 99.80 & 0.0 & \textbf{100.0} & 0.0 
              & \textbf{100.0} & 0.0 & \textbf{100.0} & 0.0 \\
            \hline

            \multirow{2}{*}{CelebA}
            & VGG11 
              & 90.40 & 0.0 & 91.10 & 0.0
              & 86.80 & 0.0 & 96.30 & 0.0 
              & \textbf{100.0} & 0.0 & \textbf{100.0} & 0.0 \\
            & ResNet50 
              & 91.60 & 0.0 & 92.30 & 0.0 
              & 87.00 & 0.0 & 82.00 & 0.0 
              & \textbf{100.0} & 0.0 & \textbf{100.0} & 0.0\\
            \hline

            \multirow{2}{*}{Skin Cancer}
            & VGG11 
              & 90.30 & 0.0 & 93.00 & 0.0  
              & 56.40 & 0.0 & 76.10 & 0.0 
              & 99.70 & 0.0 & 99.30 & 0.0 \\
            & ResNet50 
              & 91.20 & 0.0 & 92.20 & 0.0 
              & 64.00 & 0.0 & 80.30 & 0.0 
              & \textbf{100.0} & 0.0 & 99.60 & 0.0 \\
            \specialrule{1pt}{2pt}{2pt}
        \end{tabular}}
    \label{tab:exp41_featc_result}
\end{table*}

Figure~\ref{fig:exp2.1_colluding_cm} depicts the confusion matrix for average hypothesis tests’ p-values over different clients.
When the p-value exceeds 0.05, we accept the null hypothesis ($\mathcal{H}_{0}$),
  indicating that clients $u$ and $v$ are colluding adversaries. 
The observation from Figure~\ref{fig:exp2.1_colluding_cm} indicates 
  that \textsc{FDINet} achieves high p-values along the diagonal of the confusion matrix, 
  indicating its effectiveness in identifying colluding adversaries. 
However, our method doesn’t performant well in distinguishing between Knockoff and ActiveThief attacks.
This challenge arises because both attacks utilize the same surrogate dataset, 
  resulting in similar FDI for these two attacks.
Figure~\ref{fig:exp2.2_CDAcc_curve} demonstrates the effectiveness of \textsc{FDINet} 
  in detecting colluding adversaries within a large-scale MLaaS platform comprising 100 clients.
Notably, \textsc{FDINet} achieves an impressive CDAcc. of \textbf{over 91\%} for all extraction attacks.
As Figure~\ref{fig:exp2.2_CDAcc_curve} illustrates, with the increasing number of colluding adversaries,
    \textsc{FDINet} can still remain high accuracy for colluding detection.
This experiment serves as compelling evidence demonstrating the capability of 
  our method to identify colluding adversaries within a large-scale MLasS platform effectively.

\subsection{Ablation Study}
\label{sec:exp3}
To further understand how different components influence \textsc{FDINet}, 
  we carry out evaluations on two significant factors within our approach, i.e., threshold $\tau_{1}$ and batch size $bs$.

\subsubsection{Impacts of Threshold}
\label{sec:exp_threshold}
As discussed in Section~\ref{sec:threshold}, the threshold is a critical factor that affects the DAcc., and the process of selecting a suitable threshold is demanding. 
To shed light on this matter, we conduct an experiment where we employ various thresholds (ranging from 0.2 to 0.8), to observe the trend in detection accuracy.
This experiment aims to provide guidance on selecting an optimal threshold that ensures accurate detection for new datasets.

\begin{figure}[!t]
  \centering
  \includegraphics[width=0.92\linewidth]{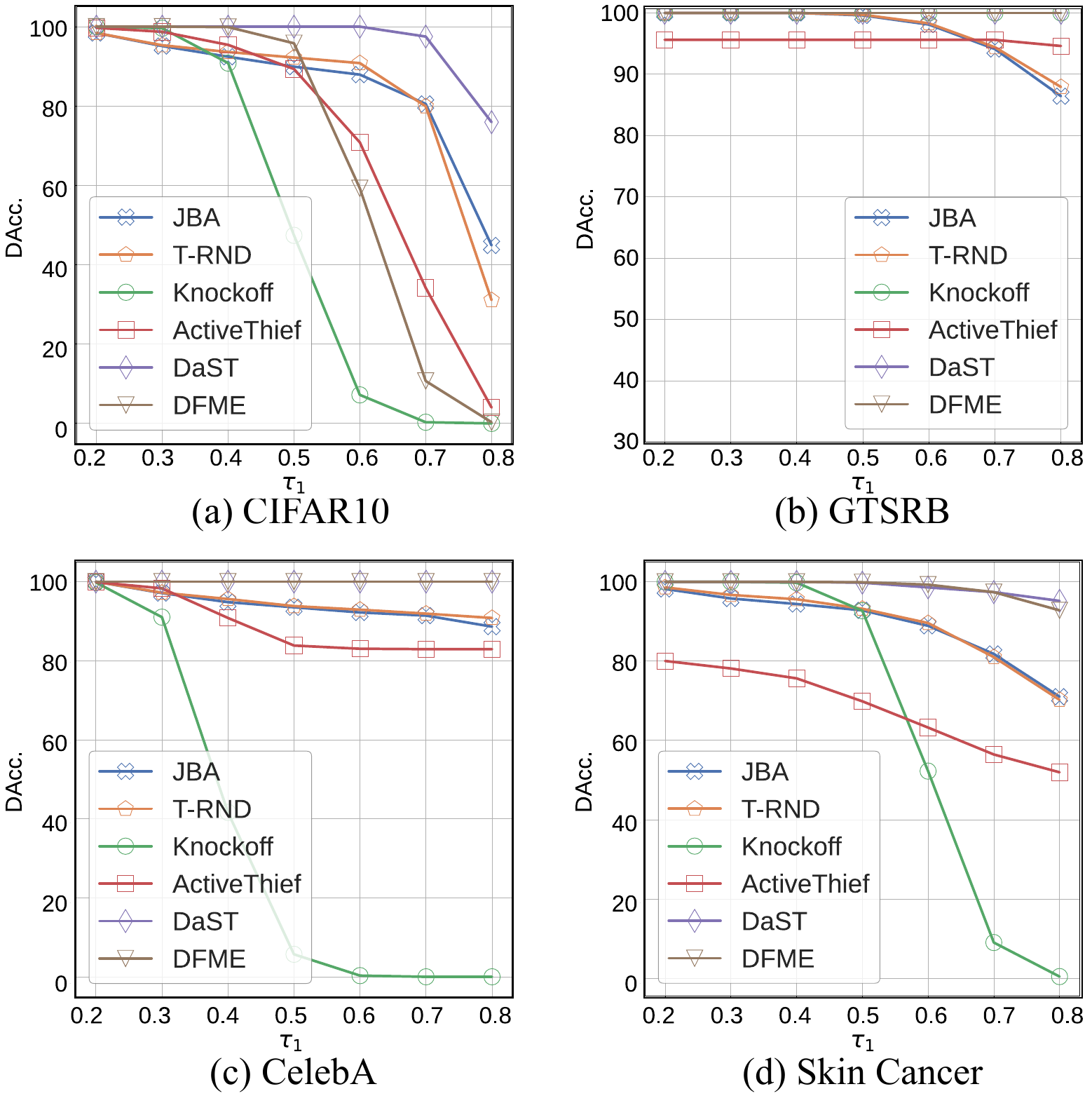}
  \caption{The performance of \textsc{FDINet} with various thresholds $\tau_{1}$.}
  \label{fig:exp31_tau1}
\end{figure}

Figure~\ref{fig:exp31_tau1} illustrates the impact of thresholds ($\tau_{1}$) on the DAcc. of \textsc{FDINet}.
It can be observed that there is a notable decrease in DAcc. as $\tau_{1}$ increases, particularly for Knockoff and ActiveThief.
This decline in accuracy can be attributed to the fact that the surrogate data used by Knockoff and ActiveThief are derived from natural images, 
  which may have a higher degree of feature overlap with $\mathcal{D}_{train}$.
In our approach, the threshold $\tau_{1}$ represents the tolerance for abnormal samples in the batch query. 
By increasing the threshold $\tau_{1}$, we can minimize false alarms since benign clients might occasionally send a few malicious queries.

\subsubsection{Impacts of Batch Size}
Relying on a single query for identifying adversaries can lead to a significant false alarms due to the limited entropy within the query's features.
To mitigate this, we adopt a majority voting strategy, as explained in Section~\ref{sec:method_detector}.
The choice of batch size ($bs$) plays a crucial role in determining the overall detection accuracy of our approach.
In order to assess the performance of our defense, we conduct an empirical evaluation 
  where we examine the effectiveness of our method across a range of batch sizes, specifically from 2 to 128.

\begin{figure}[!t]
  \centering
  \includegraphics[width=0.9\linewidth]{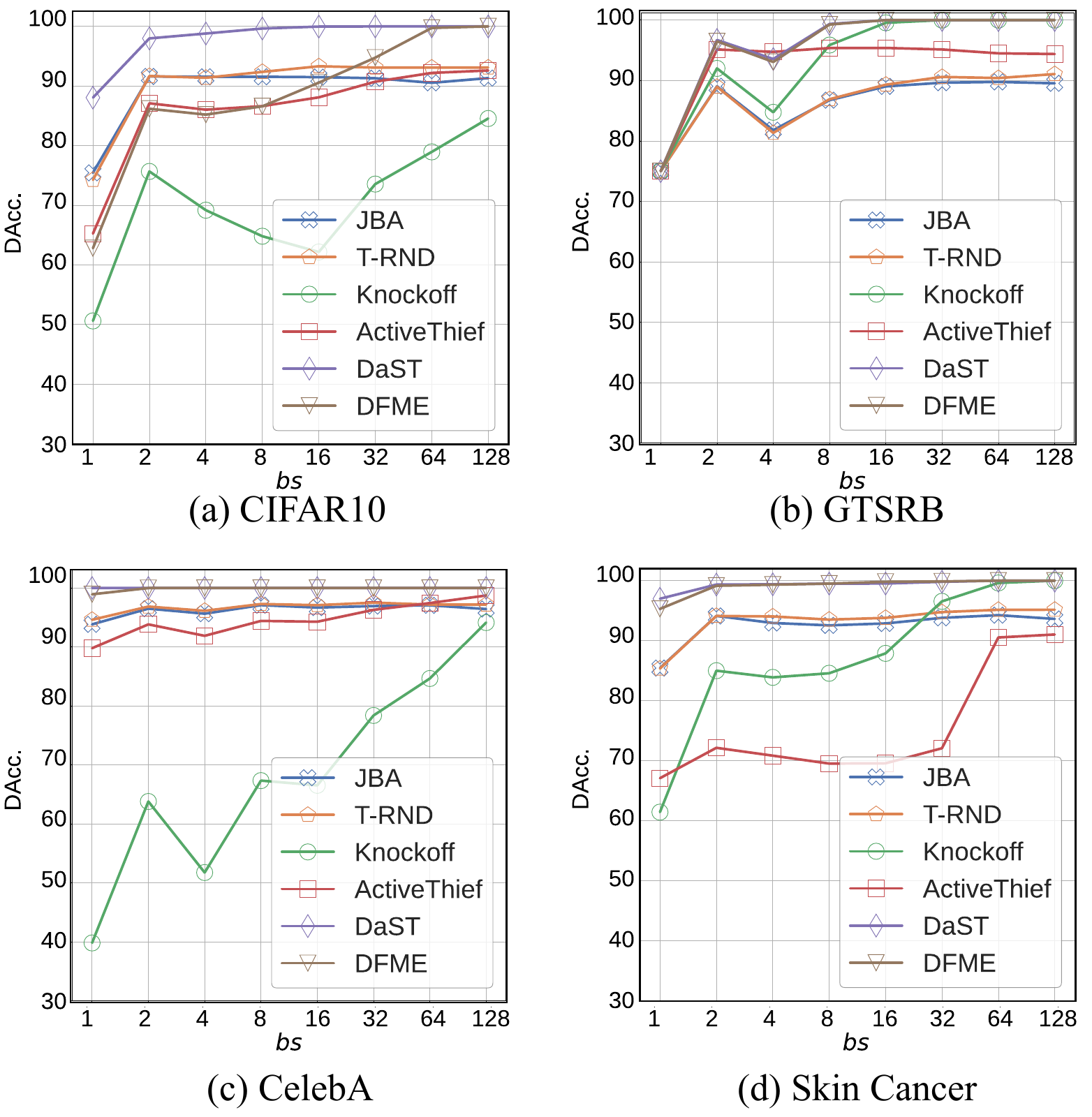}
  \caption{The performance of \textsc{FDINet} with various batch size $bs$.}
  \label{fig:exp_32_bs}
\end{figure}

Figure~\ref{fig:exp_32_bs} illustrates the impact of batch size ($bs$) on the DAcc. of \textsc{FDINet}.
It is evident from the figure that as the batch size increases, the DAcc. also improves. 
Furthermore, Figure~\ref{fig:exp_32_bs} highlights that our defense attains high DAcc. for JBA, T-RND, DFME, and DaST using just 64 queries.

\begin{table*}[!h]
    \centering
    \caption{Increased overhead to circumvent \textsc{FDINet}’s detection.}
    \resizebox{0.82\linewidth}{!}{
        \begin{tabular}{c|c|c|c|c|c|c|c}
        \specialrule{1pt}{2pt}{2pt}
            \textbf{Dataset} $\downarrow$ 
          & \textbf{Attacks}$\rightarrow$ 
          & \textbf{JBA} & \textbf{T-RND} & \textbf{Knockoff} & \textbf{ActiveThief} & \textbf{DFME} & \textbf{DaST} \\
        \specialrule{1pt}{0pt}{0pt}

        \multirow{3}{*}{CIFAR10} &
        Original queries
            & 50,000 & 50,000 & 50,000 & 50,000  &  50,000 & 50,000 \\
        &Additional queries
            & 128,924 & 116,910 & 77,732 & 96,319 & 163,709 & 89,221\\
        &Overhead
            & $+357.85\%$ & $+333.82\%$& $+255.46\%$ & $+292.64\%$ & $\textbf{+427.42\%}$ & $+278.44\%$\\
        \hline

        \multirow{3}{*}{GTSRB} &
        Original queries
            & 50,000 & 50,000 & 50,000 & 50,000  &  50,000 & 50,000 \\
        &Additional queries
            & 120,309 & 123,705 & 120,224 & 129,314 & 129,720 & 127,116 \\
        &Overhead
            & $+340.62\%$ & $+347.41\%$& $+340.45\%$ & $+358.63\%$ & $\textbf{+359.44\%}$ & $+354.23\%$\\
        \hline

        \multirow{3}{*}{CelebA} &
        Original queries
            & 50,000 & 50,000 & 50,000 & 50,000  &  50,000 & 50,000 \\
        &Additional queries
            & 200,809 & 199,703 & 76,032 & 190,312 & 219,002 & 240,634 \\
        &Overhead
            & $+501.62\%$ & $+499.41\%$& $+252.06\%$ & $+480.62\%$ & $+538.00\%$ & $\textbf{+581.27\%}$\\
        \hline

        \multirow{3}{*}{Skin Cancer} &
        Original queries
            & 50,000 & 50,000 & 50,000 & 50,000  &  50,000 & 50,000 \\
        &Additional queries
            & 90,621 & 89,133 & 75,025 & 88,424 & 99,620 & 96,019 \\
        &Overhead
            & $+281.24\%$ & $+278.27\%$& $+250.05\%$ & $+276.85\%$ & $\textbf{+299.24\%}$ & $+292.04\%$\\

        \specialrule{1pt}{2pt}{2pt}
        \end{tabular}
    }
    \label{tab:exp42_overhead}
\end{table*}

\section{Discussion}
\subsection{Adaptive Attacks}
\label{sec:exp4}
In this section, we explore two specific adaptive attacks: \textit{Feature Correction} and \textit{Dummy Query}. 
In those attacks, the adversaries know \textsc{FDINet} and potentially modify their attack strategies in order to evade our detection mechanisms.

\subsubsection{Feature Correction (FeatC)}
The adaptive adversary knows our method utilizes the FDI phenomenon to detect malicious queries. 
Consequently, the adversary can strategically modify the feature maps before submitting them to the MLaaS platform, as discussed in Section~\ref{sec:adaptive}.
In this experiment, we make two assumptions about the adversary:
(1) the adaptive adversary possesses a pre-trained encoder drawn from the model zoo (VGG11 and ResNet50), 
and (2) the adversary has access to a mini-batch of training data $\mathcal{D}_{train}$, which serves as the anchor samples.
The adaptive adversary initiates the process by generating 50,000 queries using existing model extraction attacks (such as JBA and T-RND).
Subsequently, the adversary applies the L-BFGS optimizer within \textit{FeatC} to re-correct the feature maps associated with these queries.

Table~\ref{tab:exp41_featc_result} illustrates the performance of 
    \textsc{FDINet} in defending against \textit{FeatC}, where the auxiliary encoders of \textit{FeatC} are VGG11 and ResNet50.
However, there is a slight decrease in DAcc. for Knockoff and ActiveThief, 
  as these model extraction techniques employ natural images that may overlap with the training set.
Nonetheless, \textsc{FDINet} continues to be effective in defending against the majority of attacks.

\subsubsection{Dummy Query}
PRADA introduced the \textit{Dummy Query} attack, an adaptive strategy where the adversary maintains a normal distribution of distances between queries.
Although these queries do not contribute to the surrogate model's construction, 
  they serve the purpose of evading detection.
It is assumed that the adaptive adversary possesses complete knowledge of the detection algorithm,
  including the secret detection threshold value $\tau_{1}$. 
With the objective of creating a query set comprising 50,000 samples, the adversary injects benign samples into the submitted queries.
In our evaluation, we inject a percentage $p\%$ of benign samples in the submitted queries, 
  with a batch size ($bs$) set to 50. 
We incrementally increase $p$ from 0 to 100 in intervals of 10 until the batch queries are predicted as benign by \textsc{FDINet}.
Our evaluation provides an estimated lower bound on the number of queries required to evade \textsc{FDINet} detection.

Table~\ref{tab:exp42_overhead} shows the increased overhead to circumvent \textsc{FDINet}’s detection.
The results indicate that our method increases the query overhead ranging from +252.06\% to +581.27\%.
This experiment serves as evidence that, despite the adaptive adversary's ability to distribute queries among multiple clients to evade detection, 
  our method can still enhance its query budget at least $\times 2.5$ times.

\subsection{Limitations and Future Work}
\textbf{Language model.}
This paper primarily focuses on empirical studies conducted in the field of computer vision.
However, it is crucial to recognize the significant advancements achieved in language model development.
Prominent pre-trained language transformers, including BERT and GPT-3, have been extensively employed in various downstream applications.
Nonetheless, these models are still under the threat of model extraction attacks~\cite{kalpesh2020thieves,zanella2021grey}.
We believe that our proposed method is able to transfer to language models.
In the future, we plan to extend this research to encompass language models and devise a novel model extraction detector approach specifically designed for the NLP domain.

\section{Conclusion}
\label{sec:conclusion}
This paper introduces FDI, a metric that quantitatively measures the deviation in the feature distribution of incoming queries.
Through FDI, we develop both an extraction attacks detector and a colluding adversaries detector.
Extensive experiments demonstrate the effectiveness and efficiency of \textsc{FDINet} in detecting extraction attacks.
Furthermore, \textsc{FDINet} exhibits robustness in identifying stealthiness attacks, including distributed attacks, \textit{Dummy Query}, and \textit{Feature Correction}. 
We hope this research can contribute to building a more secure MLaaS platform and promoting the scientific community's awareness of defending against model extraction attacks.

\section*{Acknowledgement}
We thank the anonymous reviewers for their feedback in improving this paper.
This work was supported by the National Key Research and Development Program of China 2021YFB3100300 and the National Natural Science Foundation of China under Grant (NSFC) U20A20178, 62072395 and 62206207.

{ 
  \small
  \bibliographystyle{IEEEtran}
  \bibliography{main}
}

\begin{IEEEbiography}[{\includegraphics[width=1in,height=1.25in,clip,keepaspectratio]{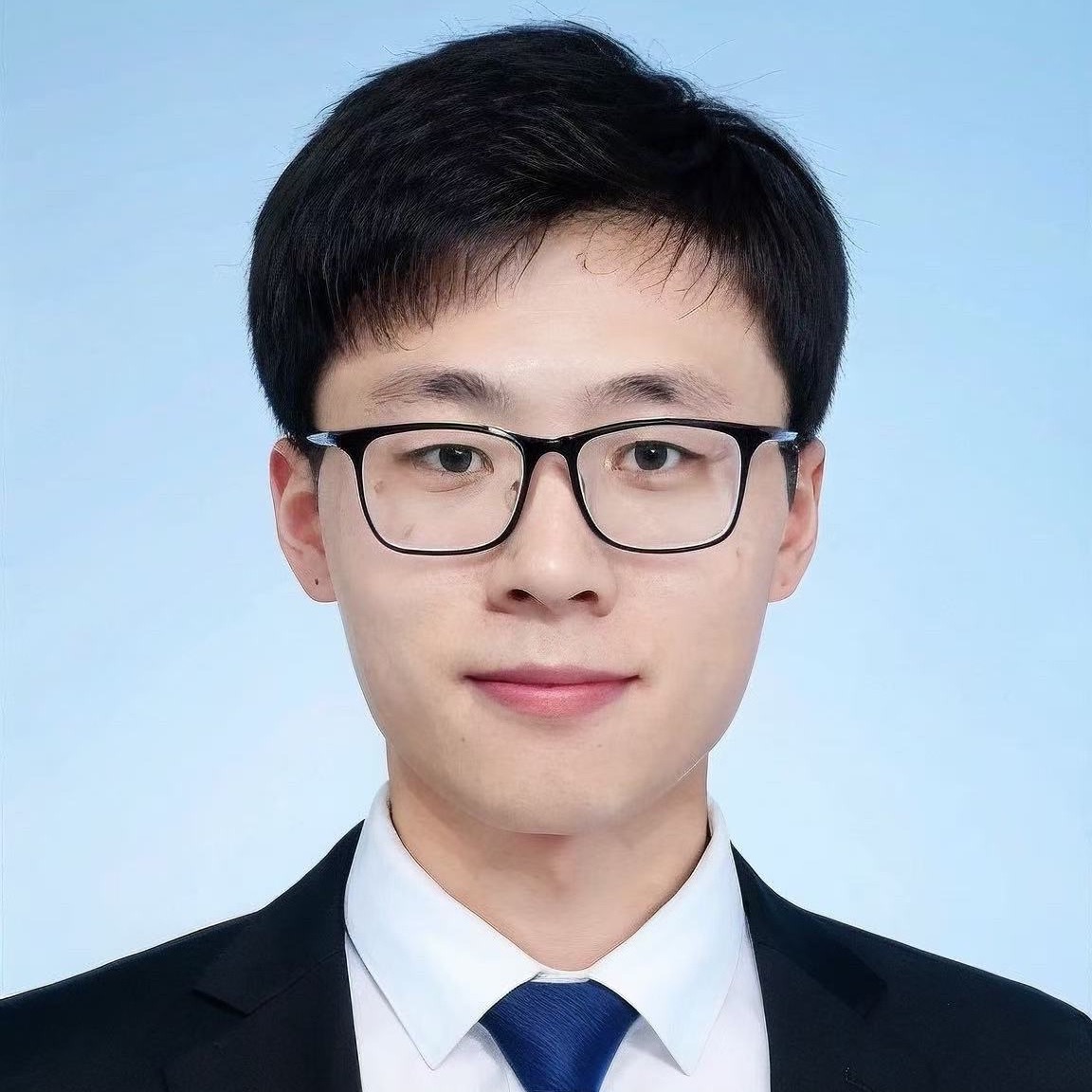}}]{Hongwei~Yao}
is currently a Postdoctoral Fellow at City University of Hong Kong working with Prof. Cong WANG.
Prior to that, he obtained his PhD (2024) degree from the College of Computer Science and Technology, Zhejiang University, advised by Dr. Zhan QIN.
His research focuses on Trustworthy AI, LLM Security and Safety.
\end{IEEEbiography}
\begin{IEEEbiography}[{\includegraphics[width=1in,height=1.25in,clip,keepaspectratio]{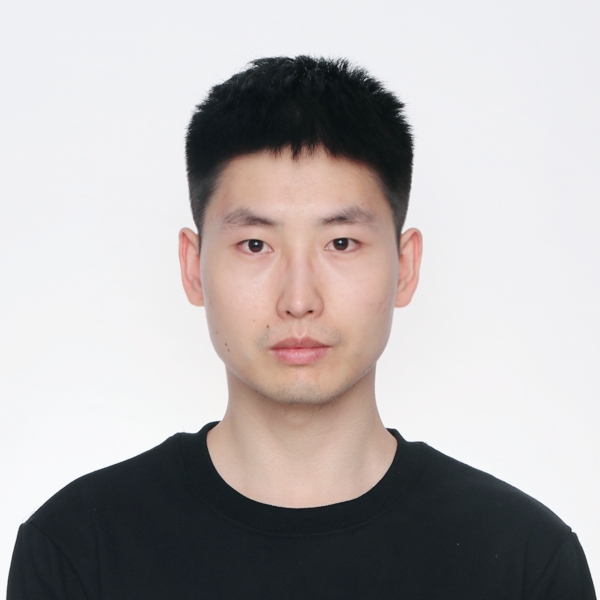}}]{Zheng~Li}
is currently a Ph.D. student at CISPA Helmholtz Center for Information Security, advised by Dr. Yang Zhang. Prior to that, he obtained his bachelor (2017) and master (2020) degrees from Shandong University under the supervision of Prof. Shanqing Guo. His research focuses on machine learning security and privacy.
\end{IEEEbiography}
\begin{IEEEbiography}[{\includegraphics[width=1in,height=1.25in,clip,keepaspectratio]{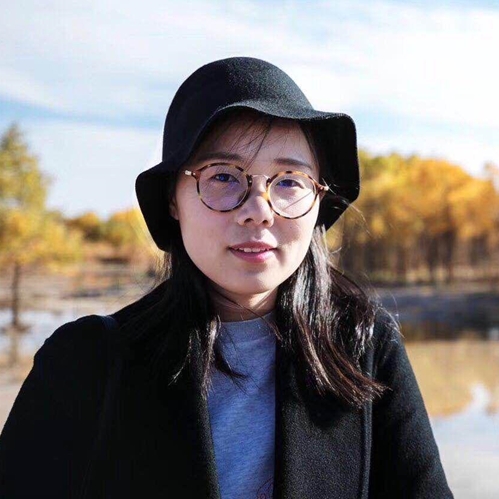}}]{Haiqin~Weng}
is currently an algorithm engineer at Ant Group.
Prior to that, she obtained her PH.D degree from the College of Computer Science and Technology, Zhejiang University, advised by Dr. Shouling Ji. 
Her research focuses on privacy preserving machine learning and secure computation outsourcing.
\end{IEEEbiography}
\begin{IEEEbiography}[{\includegraphics[width=1in,height=1.25in,clip,keepaspectratio]{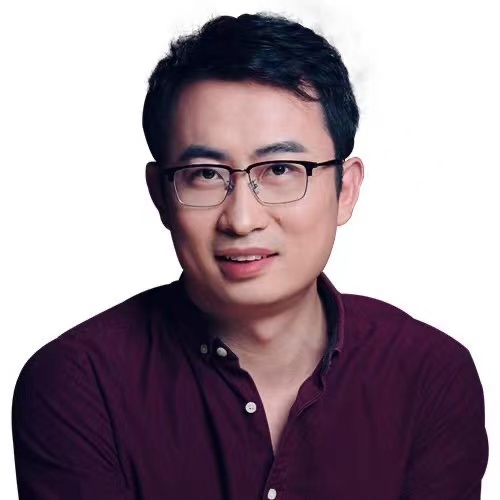}}]{Feng~Xue}
currently holds the position of Qiu Shi Innovation Scholar (Senior Level) at Zhejiang University's Hangzhou International Science and Innovation Center, and serves as Deputy Director of the Digital Security Innovation Center at the university's Jiaxing Research Institute. Previously, he was a Senior Security Expert at Alibaba Group and Ant Group, where he focused on data security, intelligent offense and defense, and AI security. As an invited member of the Chinese Artificial Intelligence Society's AI Ethics Committee, he contributes to fostering ethical practices within the AI industry. Feng Xue was also acknowledged as an advanced worker in financial big data innovation and application during the "Financial Big Data Innovation Application" case selection event, organized by the Financial Technology Professional Committee of China Payment and Clearing Association.
\end{IEEEbiography}
\begin{IEEEbiography}[{\includegraphics[width=1in,height=1.25in,clip,keepaspectratio]{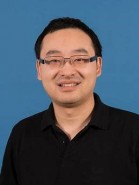}}]{Zhan Qin}
is currently a ZJU100 Young Professor, with both the College of Computer Science and Technology and the Institute of Cyberspace Research (ICSR) at Zhejiang University, China. He was an assistant professor at the Department of Electrical and Computer Engineering in the University of Texas at San Antonio after receiving the Ph.D. degree from the Computer Science and Engineering department at State University of New York at Buffalo in 2017. His current research interests include data security and privacy, secure computation outsourcing, artificial intelligence security, and cyber-physical security in the context of the Internet of Things. His works explore and develop novel security sensitive algorithms and protocols for computation and communication on the general context of Cloud and Internet devices.
\end{IEEEbiography}
\begin{IEEEbiography}[{\includegraphics[width=1in,height=1.25in,clip,keepaspectratio]{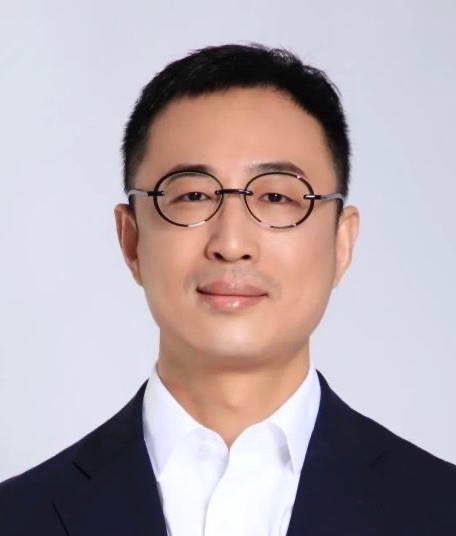}}]{Kui Ren}
is a Professor and the Dean of School of Cyber Science and Technology at Zhejiang University. Before that, he was SUNY Empire Innovation Professor at State University of New York at Buffalo. He received his PhD degree in Electrical and Computer Engineering from Worcester Polytechnic Institute. Kui’s current research interests include Data Security, IoT Security, AI Security, and Privacy. He received Guohua Distinguished Scholar Award from ZJU in 2020, IEEE CISTC Technical Recognition
Award in 2017, SUNY Chancellor’s Research Excellence Award in 2017, Sigma Xi Research Excellence Award in 2012 and NSF CAREER Award in 2011. Kui has published extensively in peer-reviewed journals and conferences and received the Test-of-time Paper Award from IEEE INFOCOM and many Best Paper Awards from IEEE and ACM including MobiSys’20, ICDCS’20, Globecom’19, ASIACCS’18, ICDCS’17, etc. His h-index is 74, and his total publication citation exceeds 32,000 according to Google Scholar. He is a frequent reviewer for funding agencies internationally and serves on the editorial boards of many IEEE and ACM journals. He currently serves as Chair of SIGSAC of ACM China. He is a Fellow of IEEE, a Fellow of ACM and a Clarivate Highly-Cited Researcher.
\end{IEEEbiography}
\end{document}